\documentstyle[prl,twocolumn,aps,epsf,eqsecnum]{revtex}
\begin{document}

\hoffset-1cm

\draft
\preprint{CU-TP-867}

\title{Multiparticle Bose-Einstein Correlations}

\author{Urs Achim Wiedemann }
\address{
Institut f\"ur Theoretische Physik, Universit\"at Regensburg,
D-93040 Regensburg, Germany\\
Department of Physics, Columbia University, New York, 
NY 10027 USA
}

\date{\today}
\maketitle

\begin{abstract}
Multiparticle symmetrization effects are contributions to the spectra
of Bose-symmetrized states which are not the product of pairwise 
correlations. Usually they are neglected in particle interferometric 
calculations which aim at determining the geometry of the boson emitting 
source from the measured momentum distributions. Based on a method 
introduced by Zajc and Pratt, we give a calculation of all multiparticle 
symmetrization effects to the one- and  two-particle momentum 
spectra for a Gaussian phase space distribution of emission points. 
Our starting point is an ensemble of $N$-particle Bose-symmetrized 
wavefunctions with specified phase space localization. In 
scenarios typical for relativistic heavy ion collisions, multiparticle 
effects steepen the slope of the one-particle spectrum for
realistic particle phase space densities by up to 20 MeV, and  
they broaden the relative momentum dependence of the two-particle 
correlations. We discuss these modifications and their 
consequences in quantitative detail. Also, we explain how multiparticle
effects modify the normalization of the two-particle correlator. The
resulting normalization conserves event probabilities, which
is not the case for the commonly used pair approximation. Finally, 
we propose a new method of calculating Bose-Einstein weights from the 
output of event generators, taking multiparticle correlations into account.
\end{abstract} 

\section{Introduction}\label{sec1}

Most hadrons are emitted in the final stage of a relativistic
heavy ion collision. They do not probe directly the hot
and dense intermediate stages where quarks and gluons are
expected to be the relevant physical degrees of freedom for
equilibration processes. The geometrical size and dynamical
state of the hadronic phase space emission region, however,
depends sensitively on the entire evolution of the 
collision. This motivates current attempts to reconstruct
its spatial and dynamical state from the experimental hadron spectra
and to use it as a starting point for a dynamical back 
extrapolation into the hot and dense intermediate stages~\cite{HW97}. 
Two-particle correlations of identical particles which are sensitive 
to the space-time characteristics of the collision~\cite{GKW79}, play 
a crucial role in this approach. The reconstruction program based on
their analysis has very
good prospects: due to the increasing event multiplicities 
and larger statistics of the CERN SPS lead beam program (and the yet
better quality data expected from Relativistic Heavy Ion Collider
RHIC at BNL), particle interferometric
measurements start showing statistical errors on the few percent level. 
Also, systematic errors are increasingly better understood. 
Theoretical calculations should aim for a similar accuracy and
control necessary approximations quantitatively.

One uncontrolled approximation used so far in almost all particle
interferometric calculations is to neglect for the 
particle momentum spectra of Bose-Einstein
symmetrized $N$-particle states all multiparticle correlations
which cannot be written in terms of simpler pairwise ones.
This reduces the number of $N!$ terms contributing to the
two-particle correlator $C({\bf K}, {\bf q})$ to a manageable sum 
over all particle pairs. Beyond this approximation,
two approaches have been used in the literature. First, Zajc has
employed Monte Carlo techniques~\cite{Z87} to generate
events with realistic multiparticle correlations. This amounts
to a shifting prescription $\lbrace {\bf p}_i\rbrace$
$ \to$ $\lbrace {\bf p}'_i\rbrace$ which modifies the
momentum distribution of simulated events according to unit weights (which
themselves depend on the space-time structure of the
source). Second, Zajc has found~\cite{Z87} the first steps towards
a calculational scheme brought into final form by Pratt~\cite{P93}:
for model distributions, the $N$-particle
spectra are given by a simple algorithm involving only two
types of terms: $C_m$ and $G_m$. In practice, this reduces the sums over
all $N!$ permutations, typical for the calculation of momentum
spectra, to sums over all partitions of $N$. 

For both these approaches, there are first numerical 
calculations~\cite{Z87,P93,Zh93} and related analytical 
attempts~\cite{BK95,AL96,FW97,W97,CZ97} to control multiparticle 
effects to the one- and two-particle spectra,
but a detailed study of their momentum dependence is missing,
even for simple models. This work aims
at filling this gap, making quantitative statements about the 
extent to which the slope of the 
one-particle spectrum and the relative momentum dependence
of the two-particle correlations are modified due to 
multiparticle symmetrization effects. Our investigation
takes the set $\Sigma$ of phase space emission 
points $({\bf p}_i,{\bf r}_i,t_i)$ as {\it initial 
condition}. For notational simplicity, we restrict the discussion 
to one particle species, like-charge pions say. To the set $\Sigma$, 
we associate a symmetrized $N$-particle wave function~\cite{Weal97}
  \begin{equation}
    \Psi_N(\vec{\bf X},t) = {1\over \sqrt{N!}}\, 
                          \sum_{s\in {\cal S}_N} 
              {\left({ \prod_{i=1}^N f_{s_i}({\bf X}_i,t) }\right)}\, ,
    \label{1.1}
  \end{equation}
where the sum runs over all permutations $s \in {\cal S}_N$ of the
$N$ indices, $\vec{\bf X}$ is a shorthand for the $N$ 3-dimensional 
coordinates ${\bf X}_i$, and the functions $f_i$ denote single 
particle wavepackets centered around $({\bf p}_i,{\bf r}_i)$ at 
initial time $t_i$ and propagated according to the free time-evolution.
Final state interactions, which imply a structure of the $N$-particle
state different from (\ref{1.1}), will not be considered in the present work.
The wave function $\Psi_N$ {\it defines} the boson emitting source. 
What we are interested in is the calculation of the one- and two-particle
momentum spectra resulting from (\ref{1.1}), the information
they contain about the initial distribution of the `emission points' 
$z_i = ({\bf p}_i,{\bf r}_i,t_i)$, the extent to which these
results modify the predictions of the pair approximation,
and finally the algorithm which implements the numerical calculation
of multiparticle spectra from the initial distribution $\Sigma$.

Our work is organized as follows: Section~\ref{sec2} shortly sets 
up and illustrates the general formalism via which particle momentum spectra 
are calculated from an $N$-particle state.
In section~\ref{sec3}, we discuss the properties of Gaussian
wavepackets which we choose for the single-particle states
$f_i({\bf X}_i,t)$ in (\ref{1.1}). 
Section~\ref{sec4} contains the main results. It gives
a complete qualitative and quantitative analysis of multiparticle
contributions to the one- and two-particle spectra for 
a boson emitting source of Gaussian phase space distribution.
In section~\ref{sec5}, we discuss the implications of this model 
study for an algorithm which calculates the Hanbury Brown Twiss
(HBT) radius parameters 
and momentum spectra for an arbitrary distribution of emission points.
Results and related conceptual questions are summarized and discussed 
in the Conclusions.

\section{The formalism}\label{sec2}

 We want to determine for the $N$-particle symmetrized 
wave function $\Psi_N$ the detection probability for measuring
$N$ identical bosons at time $t$ at the positions
${\bf X}_i$ and momenta ${\bf P}_i$. We calculate first
the $N$-particle Wigner phase space density for $\Psi_N$~\cite{HOSW84}
  \begin{mathletters}
    \label{2.1}
  \begin{eqnarray}
    &&{\cal W}_N(\vec{\bf X},\vec{\bf P},t) = (2\pi)^{3N} 
            \Psi_N(\vec{\bf X},t) 
            \nonumber \\
            &&\qquad \qquad  \, \times \left({
            \prod_{l=1}^N\, \delta^{(3)}({\bf P}_l-\hat{\bf P}_l) }\right)\, 
            \Psi_N^*(\vec{\bf X},t)
            \nonumber \\
    && \qquad \qquad = {1\over N!}\, \sum_{s,s'\in {\cal S}_n} \prod_{l=1}^N 
    W_{s'_l\, s_l}({\bf X}_l,{\bf P}_l,t)\, ,
    \label{2.1a} \\
    && W_{ij}({\bf X},{\bf P},t) = (2\pi)^3 f_{i}({\bf X},t)
            \delta^{(3)}({\bf P}-\hat{\bf P})\, f_j^*({\bf X},t)
            \nonumber \\
    &&\qquad \qquad = \int d^3{\bf y}\, {\rm e}^{-i{\bf P\cdot y}}\, 
        f_i({\bf X}+\textstyle{{\bf y}\over 2},t)
        f_j^*({\bf X}-\textstyle{{\bf y}\over 2},t)\, .
    \label{2.1b}
  \end{eqnarray}
  \end{mathletters}
Here, $\hat{\bf P}$ denotes the momentum operator acting on $\Psi_N$.
The one-particle pseudo-Wigner functions 
$W_{ij}({\bf X},{\bf P},t)$ provide the basic building blocks
for the calculation of the $N$-particle momentum spectrum which
is obtained by integrating (\ref{2.1a}) over all spatial coordinates,
  \begin{mathletters}
    \label{2.2}
  \begin{eqnarray}
    {\cal P}_N(\vec{\bf P}) &=& {\cal N}_{\Psi}\int d^3\vec{\bf X}\, 
       {\cal W}_N(\vec{\bf X},\vec{\bf P},t)
    \nonumber \\
    &=& {{\cal N}_{\Psi}\over N!}\, \sum_{s,s'\in {\cal S}_n}
    \prod_{l=1}^N {\cal F}_{s'_l\, s_l}({\bf P}_l)\, ,
    \label{2.2a} \\
    {\cal F}_{ij}({\bf P}) &=& \int d^3{\bf X}\,  
    W_{ij}({\bf X},{\bf P}) = D_i({\bf P})\, D_j^*({\bf P})\, ,
    \label{2.2b}
  \end{eqnarray}
  \end{mathletters}
where ${\cal N}_{\Psi}$ is a normalization constant, 
${\cal N}_{\Psi} = 1/ \langle \Psi_N\vert\Psi_N\rangle $,
ensuring that the probability of detecting $N$ particles
is one.
For a free time evolution of $\Psi_N$, the integration over the
spatial components of (\ref{2.2a}) leads to a time-independent
expression ${\cal P}_N(\vec{\bf P})$, since interactions between the
particles are necessary to change the momentum distribution 
in time. In contrast, integrating ${\cal W}_N(\vec{\bf X},\vec{\bf P},t)$ 
over all momenta leads to the detection probability of the $N$ bosons at
positions ${\bf X}_i$, which is a time-dependent quantity since 
free evolving bosons change their positions in time. 
The calculation of the $N$-particle momentum
spectrum according to (\ref{2.2a}) involves a sum over 
$(N!)^2$ terms. Due to the factorization of  
${\cal F}_{ij}({\bf P})$, this reduces to a sum over $N!$ terms,
  \begin{equation}
    {\cal P}_N(\vec{\bf P}) = {{\cal N}_{\Psi}\over N!}\, 
    \Bigg\vert \sum_{s\in {\cal S}_n} 
    \prod_{l=1}^N D_l({\bf P}_{s(l)}) \Bigg\vert^2\, .
    \label{2.3}
  \end{equation}
In what follows, we are especially interested in the one- and 
two-particle momentum spectra ${\cal P}^{1}_{N}({\bf P}_1)$, 
${\cal P}_{N}^{2}({\bf P}_1,{\bf P}_2)$ associated with the $N$-particle 
state $\Psi_N$. These are obtained by integrating ${\cal P}_N(\vec{\bf P})$ 
over all but one, respectively two momenta,
  \begin{mathletters}
    \label{2.4}
  \begin{eqnarray}
    {\cal P}^1_N({\bf P}_1) &=& 
    {{\cal N}_{\Psi}\over N!} \sum_{s,s'} {\cal F}_{s'_1\, s_1}({\bf P}_1)
    \prod_{l=2}^N f_{s'_l\, s_l}\, , 
    \label{2.4a} \\
    {\cal P}^2_N({\bf P}_1,{\bf P}_2) &=& 
    {{\cal N}_{\Psi}\over N!} \sum_{s,s'} {\cal F}_{s_1's_1}({\bf P}_1)\, 
    {\cal F}_{s_2's_2}({\bf P}_2)
    \prod_{l=3}^N f_{s_l's_l}\, ,
    \label{2.4b} \\
    f_{ij} &=& \int d^3{\bf P}\, {\cal F}_{ij}({\bf P})\, .
    \label{2.4c}
  \end{eqnarray}
  \end{mathletters}
All particle momentum spectra are given in terms of the
building blocks $D_i({\bf P})$ (which determine 
${\cal F}_{ij}$) and $f_{ij}$. Once the analytical
form of the single particle wavefunctions $f_i$ is specified, these
are readily calculated. In what follows, capital letters
denote measurable position and momentum coordinates, small letters
${\bf p}_i$, ${\bf r}_i$, $t_i$ denote the centers of wavepackets 
which are not directly measurable. The only exception to this
is the measurable relative momentum ${\bf q} = {\bf P}_1 - {\bf P}_2$
of the two-particle correlator $C({\bf K} = \textstyle{1\over 2}
({\bf P}_1 + {\bf P}_2), {\bf q})$ which we denote by a small
letter.

\subsection{The Zajc-Pratt algorithm}\label{sec2a}

Dynamical correlations between particles in the source are reflected
in correlations in the set of emission points $({\bf p}_i,{\bf r}_i,t_i)$.  
If there are no correlations, then the initial distribution of the centers 
of single particle wavepackets is given by a one-particle
probability distribution $\rho({\bf p},{\bf r},t)$. The
$n$-particle spectra for a set of events with multiplicity
$N$ are obtained by averaging over this distribution
  \begin{mathletters}
    \label{2.5}
  \begin{eqnarray}
    \overline{\cal P}_N^{n}({\bf P}_1,...,{\bf P}_n)
    &=& \int {\left({ \prod_{i=1}^N {\cal D}\rho_i }\right)} \,
        {\cal P}_N^{n}({\bf P}_1,...,{\bf P}_n)\, ,
        \label{2.5a}\\
    {\cal D}\rho_i &=& d^3{\bf r}_i\, d^3{\bf p}_i\, dt_i\,
       \rho({\bf p}_i,{\bf r}_i,t_i)\, .
       \label{2.5b}
  \end{eqnarray}
  \end{mathletters}
A particular model distribution with correlations which
assumes that the emission probability of a boson is 
increased if there is another emission in its vicinity~\cite{CZ97},
is e.g. obtained by replacing in (\ref{2.5a})
  \begin{equation}
    \prod_{i=1}^N \rho({\bf p}_i,{\bf r}_i,t_i)
    \longrightarrow 
    \prod_{i=1}^N \rho({\bf p}_i,{\bf r}_i,t_i)
    {\langle\Psi_N\vert\Psi_N\rangle \over w(N)}\, .
    \label{2.5ext}
  \end{equation}
Here $\omega(N)$ is an averaged normalization defined below.
The technical advantage of adopting (\ref{2.5ext}) is that the 
$({\bf p}_i,{\bf r}_i, t_i)$-dependent normalization factor
${\cal N}_{\Psi}$ in the spectrum ${\cal P}_N^n$ of (\ref{2.5a})
is canceled. This allows to write without approximation all
spectra in terms of the building blocks~\cite{P93}
  \begin{mathletters}
    \label{2.6}
  \begin{eqnarray}
    G_m({\bf P}_1,{\bf P}_2) &=& \int {\left({ \prod_{l=1}^m
    {\cal D}\rho_{i_l} }\right)}\, D_{i_1}^*({\bf P}_1) 
    f_{i_1i_2}\, f_{i_2i_3} \times ...
    \nonumber \\
    && \qquad \qquad \times f_{i_{m-1}i_m} D_{i_m}({\bf P}_2)\, ,
    \label{2.6a} \\
    C_m &=& \int d^3{\bf P} G_m({\bf P},{\bf P})\, .
    \label{2.6b}
  \end{eqnarray}
  \end{mathletters}
The resulting Zajc-Pratt (ZP) algorithm for the calculation of one- and
two-particle spectra reads~\cite{Z87,P93,CZ97}
  \begin{eqnarray}
    &&w(N) = \sum_{(n,l_n)_N} 
    {N! \over {\prod_n n^{l_n} (l_n!)}}
    C_1^{l_1}\, C_2^{l_2}\, ...\, C_n^{l_n}\, , 
    \label{2.7} \\
    &&\overline{\cal P}_N^{1}({\bf P}) = 
    \sum_{m=1}^N {{(N-1)!}\over {(N-m)!}}\, {w(N-m)\over w(N)}\, 
    G_m({\bf P},{\bf P})\, ,
    \label{2.8} \\
    &&\overline{\cal P}_N^{2}({\bf P_1},{\bf P_2}) = 
    \sum_{J=2}^N {{(N-2)!}\over {(N-J)!}}\, {w(N-J)\over w(N)}\,
    \nonumber \\ 
    && \qquad \times \sum_{i=1}^{J-1} 
    \Big\lbrack
         G_i({\bf P}_1,{\bf P}_1)G_{J-i}({\bf P}_2,{\bf P}_2)
    \nonumber \\
    && \qquad \qquad \, 
    + G_i({\bf P}_1,{\bf P}_2)G_{J-i}({\bf P}_2,{\bf P}_1)
    \Big\rbrack \, .
    \label{2.9}
  \end{eqnarray}
These spectra are normalized to unity. 
In Appendix~\ref{appb}, we give their derivation 
in some combinatorical detail. 
Due to the ZP-algorithm, the calculation of the $N$-particle spectra
is reduced from sums over all permutations to sums over all
partitions $(n,l_n)_N$ of a set of 
$N$ points into $l_n$ subsets of $n$ points,  
$N = \sum_n l_n n $. The number of partitions of $N$ 
grows asymptotically like $e^{\sqrt{N}}$. For explicit
calculations with event multiplicities in the hundreds, it
is hence important to get control over the
$m$-dependence of the Pratt terms $C_m$ and $G_m$. This is our
strategy in section~\ref{sec4}.  
%
\subsection{A simple example: the Zajc model}\label{sec2b}

To illustrate the above formalism, we consider a normalized 
$N$-particle density matrix $\rho_{\pi}^{(N)}$ for multiparticle states
$|{\bf x}_1, ...,{\bf x}_N\rangle$, created by repeated operation
of the single-particle creation operator 
$\phi^\dagger({\bf x})$,
  \begin{mathletters}
    \label{2.10}
  \begin{eqnarray}
    \phi^\dagger({\bf x}) &=& \int 
    {d^3{\bf k}\over (2\pi)^{\footnotesize{3/2}}} 
    {\rm e}^{i{\bf k}\cdot {\bf x}}\, g({\bf k})\, a_{\bf k}^\dagger\, ,
    \label{2.10a} \\
    \rho_\pi^{(N)} &=& {\cal N} \int d{\bf x}_1 ... d{\bf x}_N
    \rho({\bf x}_1) ... \rho({\bf x}_N) 
    \nonumber \\
    && \times \, |{\bf x}_1, ...,{\bf x}_N\rangle
    \langle{\bf x}_1, ...,{\bf x}_N|\, .
    \label{2.10b}
  \end{eqnarray}
  \end{mathletters}
This density matrix specifies in particular the 
one- and two-particle spectra 
${\rm Tr}[\rho_\pi^{(n)} a_{\bf P}^\dagger a_{\bf P}]$ and
${\rm Tr}[\rho_\pi^{(n)} a_{{\bf P}_1}^\dagger a_{{\bf P}_2}^\dagger$ 
$a_{{\bf P}_1} a_{{\bf P}_2}]$. 
For the Gaussian model distribution~\cite{Z87}
  \begin{mathletters}
    \label{2.11}
  \begin{eqnarray}
    { {|g({\bf k})|^2}\over (2\pi)^3}
    &=& (2\pi p_0^2)^{\footnotesize{-3/2}}
    \exp\left[- {\bf k}^2/ 2p_0^2 \right]\, ,
    \label{2.11a} \\
    \rho({\bf x}) &=& (\pi R^2)^{\footnotesize{-3/2}}
    \exp\left[- {\bf x}^2/ R^2 \right] \, ,
    \label{2.11b}
  \end{eqnarray}
  \end{mathletters}
the effects of multiparticle correlations on the HBT-radius 
parameters have been considered already by Zajc. His discussion 
however is restricted to an explicit calculation of three-particle
symmetrization effects and to qualitative estimates of higher
order contributions. Here, we demonstrate that the ZP-algorithm
allows for a complete quantitative analysis of the Zajc model.
The first step is to identify
the building blocks of the particle spectra (\ref{2.4}), 
  \begin{mathletters}
    \label{2.12}
  \begin{eqnarray}
    D_i({\bf P}) &=& {g({\bf P})\over (2\pi)^{\footnotesize{3/ 2}}}
                     \exp\left[{i{\bf P}\cdot{\bf x}_i}\right]\, ,
    \label{2.12a} \\
    f_{ij} &=& \exp\left[
      {- {p_0^2\over 2}({\bf x}_i - {\bf x}_j)^2}\right]\, .
    \label{2.12b}
  \end{eqnarray}
  \end{mathletters}
The calculation of the terms $C_m$ reduces then to 
$m$-dimensional Gaussian integrations. One finds $C_1 = 1$ and
  \begin{mathletters}
    \label{2.13}
  \begin{eqnarray}
    C_m &=& (1 + p_0^2R^2)^{-\footnotesize{3(m-1)/2}}
            \left({ {1-h}\over {\det B_m}}\right)^{\footnotesize{3/ 2}}
            \label{2.13a} \\
    \det B_m &=& {h^m\over 2^{m-1}}\, 
    \left( T_m\left(\textstyle{1\over h}\right) 
                                 -1\right)\, ,
            \nonumber \\
    h &=& {1\over 1 + {1/c}}\, , \qquad 
          c =  R^2p_0^2\, ,
          \label{2.13b}
  \end{eqnarray}
  \end{mathletters}
where $T_m$ denotes Chebysheff polynomials of the first kind~\cite{math}.
The momentum-dependent terms read
  \begin{mathletters}
    \label{2.14}
  \begin{eqnarray}
    G_m({\bf K}, {\bf q}) &=& 
    C_m\, { (1 + g_K^{(m)})^{\footnotesize{3/2}}\over
            (2\pi p_0^2)^{\footnotesize{3/2}}}\, 
          \exp\lbrack{ -{\bf K}^2\, g_K^{(m)}/2p_0^2 }\rbrack 
          \nonumber \\
          && \times 
          \exp\lbrack{ -\left( R^2 g_Q^{(m)}/4 + 1/8p_0^2\right) {\bf q}^2
            }\rbrack
          \label{2.14a} \\
   g_Q^{(m)} &=& \textstyle{1\over 4} {\vec{e}_+}^t\, A_m^{-1}\, \vec{e}_+\, ,
          \qquad e_+^{(i)} = \delta_{i1} + \delta_{im}\, ,
          \label{2.14b} \\
   g_K^{(m)} &=& \textstyle{c\over 2}\, 
   {\vec{e}_-}^t\, A_m^{-1}\, \vec{e}_- + 1\, , \, \, \, 
          e_-^{(i)} = \delta_{i1} - \delta_{im}\, ,
          \label{2.14c} \\
   \left( A_m\right)_{ij} &=& (1+c)\delta_{ij}
          - \textstyle{c\over 2}
          \left( \delta_{i,j+1} +  \delta_{i+1,j} \right)\, .
          \label{2.14d}
  \end{eqnarray}
  \end{mathletters}
The main message of these involved but explicit expressions is 
contained in the m-dependence of the terms $g_Q^{(m)}$ and $g_K^{(m)}$.
These specify the ${\bf K}$- and ${\bf q}$-dependence of the building
blocks $G_m$ and hence of all spectra. Here, they are functions of the
phase space volume $V = p_0^3\, R^3$ only, and their behaviour can
be understood by simple arguments: 
%
\begin{figure}\epsfxsize=8cm 
\centerline{\epsfbox{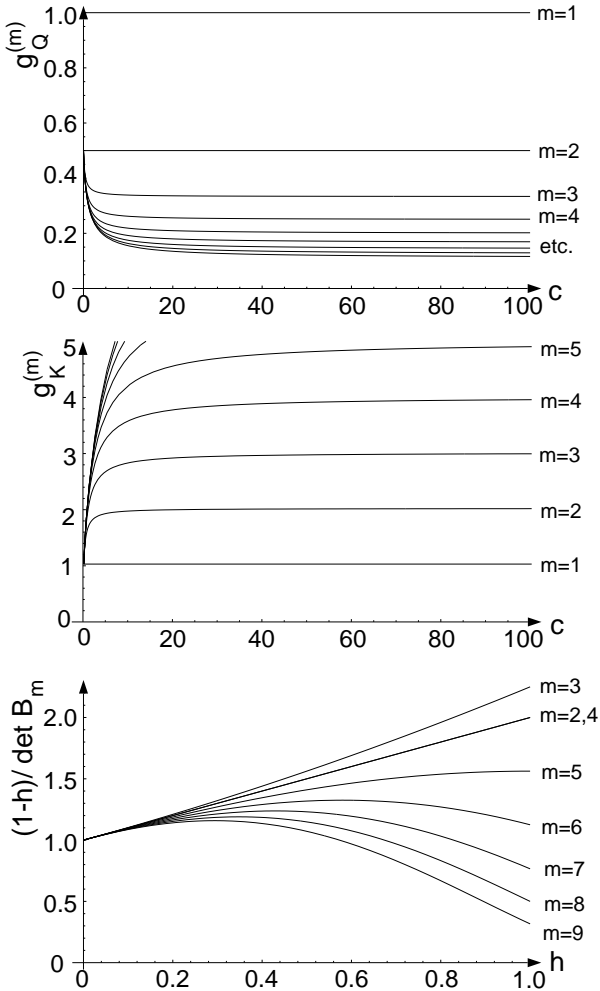}}
\caption{\it The terms $g_Q^{(m)}$, $g_K^{(m)}$ and the remainder
term $(1-h)/\det B_m$ characterize the ${\bf q}$-dependence,
${\bf K}$-dependence and weight of Pratt terms and hence
the momentum dependence of particle spectra. Their $m$-dependence
contains information about how higher order multiparticle correlations 
affect the spectra. Results are shown for the Zajc model (\ref{2.11}).
}\label{fig1}
\end{figure}
%
The factors $g_K^{(m)}$ generically increase with increasing $m$,
see Figure~\ref{fig1}. Especially, in a limiting case, one finds
  \begin{equation}
    \lim_{V\to\infty} g_K^{(m)} = m\, .
    \label{2.15}
  \end{equation}
The reason is that Bose-Einstein symmetrization effects enhance the low
momentum region of the one-particle spectrum, leading to steeper
slopes. This one-particle spectrum $\overline{\cal P}_N^{1}({\bf P})$ 
is a linear superposition of Gaussian terms $G_m({\bf K}, {\bf q} = 0)$,
which due to $g_K^{(m)}$ show increasingly steeper slopes.

 The ${\bf q}$-dependence of $G_m$ governs the ${\bf q}$-dependence
of the two-particle spectrum. The terms $g_Q^{(m)}$ depicted in
Figure~\ref{fig1} decrease with increasing $m$ and have the 
limiting values
  \begin{equation}
    \lim_{V\to\infty} g_Q^{(m)} = {1\over m}\, .
    \label{2.16}
  \end{equation}
Zajc has concluded on the basis of this behaviour 
that~\cite{Z87} ``the two-particle correlation function becomes a 
superposition of terms with successively broader distribution in 
${\bf P}_1 - {\bf P}_2$, leading to an increasingly smaller value
for the inferred radius.'' The reason is that Bose-Einstein symmetrization
effects enhance the probability of finding bosons closer together
in configuration space and hence result in broader ${\bf q}$-distributions
of the two-particle spectra.

The above arguments explain the effect of multiparticle 
correlations qualitatively. For a quantitative 
understanding, the weights of higher
order terms contributing to the one- and two-particle
spectra are important. These weights are
governed by the terms $C_m$ which (up to a correction factor
$(1-h)/\det B_m$ of order unity) essentially decrease like 
$(m-1)$th powers of the inverse phase space volume.
For event multiplicities in the hundreds, a
quantitative analysis can then be done
numerically, using the analytical expressions (\ref{2.13})
and (\ref{2.14}). We defer such a study to a slightly more general model in 
section~\ref{sec4} where certain analytical simplifications
allow for a more transparent discussion. A short comparison
of the qualitative and quantitative properties of the models
studied here and in section~\ref{sec4} is given in the text
following Eq.~(\ref{4.4}).

\section{Gaussian wavepackets}\label{sec3}

In the example of section~\ref{sec2b},
we have started from single particle creation operators
$\phi^{\dagger}$ whose momentum support $g({\bf k})$ does 
not carry a label $i$. As a consequence, the $N$-particle states 
considered in section~\ref{sec2b} are build up from $N$ single particle
wavefunctions with {\it identical} phase space localization.
We now adopt a more general setting in which a set of $N$
phase space points $({\bf p}_i, {\bf r}_i, t_i)$ is associated with
$N$ Gaussian wavepackets $f_i$, centered at initial time $t_i$ 
around the points $({\bf p}_i,{\bf r}_i)$ with spatial width 
$\sigma$,~\cite{PGG90,MP97,Weal97}
  \begin{equation}
    f_i({\bf X},t_i) 
    = {1\over (\pi\sigma^2)^{\footnotesize{3/4}} }
    \exp{\Big\lbrack{-\textstyle{1\over 2\sigma^2}\, 
      ({\bf X}-{\bf r}_i)^2 + i{\bf p}_i\cdot {\bf X} }\Big\rbrack}\, .
    \label{3.1}
  \end{equation}
The Fourier transform of $f_i$ is proportional to $\exp \left[
- \sigma^2({\bf k}-{\bf p}_i)^2/2 \right.$ 
$\left. - i{\bf r}_i\cdot ({\bf k}-{\bf p}_i)\right]$ and can 
be compared to the momentum support of $\phi^{\dagger}$ in (\ref{2.10a}). 
The corresponding $N$-particle symmetrized states reduce to those
considered in section~\ref{sec2b} for $p_0 = 1/\sigma$ and
${\bf p}_i = 0 = {\bf r}_i$, when the momentum support function
becomes independent of the particle label $i$.

The one boson state (\ref{3.1}) is optimally
localized around $({\bf p}_i,{\bf r}_i)$ in the sense that it saturates the
Heisenberg uncertainty relation $\Delta {\bf x}\cdot \Delta {\bf p}_x = 1$,
with $\Delta x_i = \sigma$ at initial time $t=t_i$. Different choices
of single particle wavefunctions can lead to different
results for the calculated spectra. We have some control over the 
extent to which different choices matter by finally checking the 
$\sigma$-dependence of our results. The model parameter can range
between $\sigma \in [0,\infty]$. In Refs.~\cite{Weal97,ZWSH97}, 
it was argued that a realistic value for $\sigma$ is for pions 
of the order of the Compton wavelength. 

The free time evolution of $f_i$ is specified by the one-particle 
hamiltonian $H_0$ which acts as a multiplication operator in 
momentum space,
  \begin{eqnarray}
    f_{z_i}({\bf X},t) 
    &=& \left( {\rm e}^{-iH_0(t-t_i)}f_i\right)({\bf X},t)
    \nonumber \\
    &=& {1\over (2\pi)^3} \int d^3{\bf k}\,
      {\rm e}^{i{\bf k}\cdot{\bf X} - iE_{\bf k}(t-t_i)}
      \tilde{f}_i({\bf k})\, .
      \label{3.2}
  \end{eqnarray}
The resulting building blocks for the $N$-particle
spectra are
  \begin{mathletters}
    \label{3.3}
  \begin{eqnarray}
     D_i({\bf P}) &=& 2^{\footnotesize{3/2}}\, 
        (\pi\sigma^2)^{\footnotesize{3/4}}
        e^{-(\sigma^2/2)({\bf p}_i-{\bf P})^2} 
        e^{iE_Pt_i - i{\bf P}\cdot{\bf r}_i}\, ,
        \label{3.3a}\\
      s_i({\bf P}) &=& {\cal F}_{ii}({\bf P}) =
         2^3\, (\pi\sigma^2)^{\footnotesize{3/2}}\, 
         {\rm e}^{- \sigma^2 
         {\left({ {\bf p}_i - {\bf P}}\right)}^2}\, .
        \label{3.3b}
  \end{eqnarray}
  \end{mathletters}
To streamline our notation, we have neglected in $D_i$ 
a factor $\exp [i{\bf r}_i\cdot {\bf p}_i]$. 
The product of these factors cancels in the calculation of 
the Wigner function (\ref{2.1b}) and a fortiori in all the 
functions derived from it. The function 
$s_i({\bf P}) = {\cal F}_{ii}({\bf P})$ 
denotes the probability that a boson in the state $f_i$ is 
detected with momentum ${\bf P}$. This measured
momentum ${\bf P}$ has a Gaussian distribution around
the central momentum ${\bf p}_i$ of the wavepacket. 

The functions $f_{ij}$ in (\ref{2.4c}) characterize the
overlap between the wavepackets $f_i$ and $f_j$ and play
an important role in the ZP-algorithm. They take a 
particularly simple form if all particles are
emitted in a flash,
  \begin{eqnarray}
    f_{ij} &\propto& \exp\left[ - {1\over 4\sigma^2}
                            ({\bf r}_i-{\bf r}_j)^2
                             - {\sigma^2\over 4}
                             ({\bf p}_i-{\bf p}_j)^2 \right]
               \nonumber \\
               && \times \exp \left[ -{i \over 2}
                    ({\bf p}_i+{\bf p}_j)\cdot
                    ({\bf r}_i-{\bf r}_j)\right]\, .
    \label{3.4}
  \end{eqnarray}
All terms contributing to ${\cal P}_N^1$ or ${\cal P}_N^2$ contain
the same number of factors $f_{ij}$ and hence, the normalization
of $f_{ij}$ does not matter in what follows. For notational
convenience, we change it by fixing $f_{ii} = 1$. The functions
$f_{ij}$ measure the distance $\vert z_i-z_j\vert$ between the
phase space points $i$ and $j$. This distance measure depends
on the wavepacket width $\sigma$ but leaves the phase space
volume independent of $\sigma$,
  \begin{eqnarray}
    \vert f_{ij}\vert &=&  
    \exp \left[{-\textstyle{1\over 4} \vert z_i-z_j\vert^2}\right]\, , 
    \nonumber \\ 
    z_j &=& {1\over \sigma}r_j + i\sigma p_j\, .
    \label{3.5}
  \end{eqnarray}
%

\section{Multiparticle correlations for a Gaussian Model}\label{sec4}

We now determine quantitatively multiparticle correlation
effects for a source of $N$ identical bosons whose wavepackets 
of spatial width $\sigma$ are emitted instantaneously according 
to a Gaussian $N$-particle phase space distribution (\ref{2.5ext})
with
  \begin{equation}
    \rho({\bf p},{\bf r}) = {\delta(t)\over \pi^3R^3\Delta^3}\, \, 
               \exp\left[ -\frac{{\bf r}^2}{R^2} 
                         -\frac{{\bf p}^2}{\Delta^2} \right]\, .
    \label{4.1}
  \end{equation}
Our main aim is to study for this model the extent to which
multiparticle correlations modify the slope of the one-particle 
spectrum and the width of the two-particle correlator. To this
end, we calculate first the building blocks of the ZP-algorithm.
Having explicit expressions for $g_Q^{(m)}$, $g_K^{(m)}$ 
and $C_m$ in terms of simple polynomials will 
simplify our discussion considerably.
The corresponding first order Pratt terms are
  \begin{mathletters}
    \label{4.2}
  \begin{eqnarray}
    C_1 &=& 1\, ,
    \label{4.2a} \\
    G_1({\bf P}_1,{\bf P}_2) &=& 
    {\left(1 + \sigma^2\Delta^2\right)}^{-\footnotesize{3/2}}
    \nonumber \\
    && \times
    \exp \Big\lbrack 
    -\left(\frac{\sigma^2}{4} + \frac{R^2}{4} \right){\bf q}^2
    \Big\rbrack
    \nonumber \\
    && \times \exp \Big\lbrack 
    -{1\over {\Delta^2 + 1/\sigma^2}} {\bf K}^2 \Big\rbrack \, .
    \label{4.2b}
  \end{eqnarray}
  \end{mathletters}
All higher order Pratt terms can be calculated explicitly as averages
over relative and average pair distributions. Details are given in
Appendix~\ref{appa}. The momentum-independent terms read
  \begin{mathletters}
    \label{4.3}
  \begin{eqnarray}
    C_m &=& (h_1^{(m)} h_2^{(m)})^{\footnotesize{-3/2}}
    \nonumber \\
    && \times 
    \left( \lbrack 1 + \textstyle{\sigma^2\Delta^2\over 2}\rbrack
    \lbrack1 + \textstyle{R^2\over 2\sigma^2}\rbrack 
    \right)^{\footnotesize{-3(m-1)/2}}\, ,
    \label{4.3a} \\
    h_1^{(m)} &=& \sum_{k=0}^{m-1} 
    \left(\begin{array}{c} m-1 \\ k\end{array}\right)
    a^{\textstyle{k\over 2}\Big\vert^h}\,\, 
    b^{\textstyle{k\over 2}\Big\vert^l}\, ,
    \label{4.3b} \\
    h_2^{(m)} &=&  
    \sum_{k=0}^{m-1} 
    \left(\begin{array}{c} m-1 \\ k\end{array}\right)
    a^{\textstyle{k\over 2}\Big\vert^l}\,\, 
    b^{\textstyle{k\over 2}\Big\vert^h}\, ,
    \label{4.3c} \\
    a &=& {1\over 1 + {2\sigma^2/R^2}}\, ,
    \qquad
    b = {1\over 1 + {2/\sigma^2\Delta^2}}\, .
    \label{4.3d}
  \end{eqnarray}
  \end{mathletters}
Here, $\textstyle{k\over 2}\vert^l$ denotes the greatest
integer not larger than $\textstyle{k\over 2}$ (`floor'), and
$\textstyle{k\over 2}\vert^h$ the least integer not smaller
than $\textstyle{k\over 2}$ (`ceiling').
The notational shorthands $a$ and $b$ range between
0 and 1 depending on the phase space localization of the 
wavepacket centers, i.e., they map the whole parameter space
$0 < R < \infty$, $0 < \Delta < \infty$ of the model (\ref{4.1})
onto a finite region. The momentum-dependent terms are given by
  \begin{mathletters}
    \label{4.4}
  \begin{eqnarray}
    &&G_m({\bf P}_1,{\bf P}_2) = C_m\, 
    \left({2\pi\over \Delta^2} b\,g_K^{(m)}\right)^{\footnotesize{3/ 2}}\,
    \nonumber \\
    && \qquad \qquad \times
    \exp \Big\lbrack{ -\left(\textstyle{\sigma^2\over 4} + 
                              \textstyle{R^2\over 8} \right)
      g_Q^{(m)}\, {\bf q}^2} \Big\rbrack
    \nonumber \\
    &&  \qquad \qquad \times
    \exp \Big\lbrack{ -\textstyle{2\over \Delta^2 }b\, g_K^{(m)}\, 
      {\bf K}^2}\Big\rbrack \, ,
   \label{4.4a}\\
   &&g_Q^{(m)} = h_3^{(m)}/ h_2^{(m)}\, ,
    \label{4.4b}\\
   && g_K^{(m)} = h_1^{(m)}/ h_3^{(m)}\, ,
    \label{4.4c} \\
   &&h_3^{(m)} = 1 + \sum_{k=1} (ab)^k\, 
       \left(\begin{array}{c} m \\ 2k \end{array}\right) \, .
    \label{4.4d} 
  \end{eqnarray} 
  \end{mathletters}
The comparison of the present model calculation with the Zajc model
(\ref{2.11}) is not straightforward. As mentioned in the sequel
of Eq. (\ref{3.1}), the wavepackets used in both models can be
compared by setting the wavepacket centers ${\bf p}_i = 0 = {\bf r}_i$.
However, the integral over $\rho({\bf x})$ in (\ref{2.10b}) performs 
an average over the positions ${\bf x}$ while in the model (\ref{4.1})
we do not average over the position ${\bf X}$ but over
the centers of wavepackets ${\bf p}_i$, ${\bf r}_i$.
Due to these different starting points, the Zajc model is not a simple
limiting case of (\ref{4.1}). Nevertheless, main features of the
Zajc model (\ref{2.11}) can be reproduced qualitatively in the present model.
The leading contribution of the momentum-independent
Pratt terms $C_m$ shows again a power law behaviour. Also,
the $m$-dependence of the terms $g_Q^{(m)}$ and $g_K^{(m)}$ 
of the Zajc model is recovered in certain limiting cases,
  \begin{mathletters}
    \label{4.5}
  \begin{eqnarray}
    \lim_{a\to 0} \lim_{b\to 1} g_Q^{(m)} &=& {1\over m}\, ,
    \label{4.5a} \\
    \lim_{a\to 1} \lim_{b\to 0} g_K^{(m)} &=& m\, .
    \label{4.5b} 
  \end{eqnarray}
  \end{mathletters}
In general, however, the $m$-dependence of the terms 
$g_Q^{(m)}$ and $g_K^{(m)}$ is much weaker in the present
model. Especially, there is no limit in which both
$g_Q^{(m)} = \textstyle{1\over m}$ and $g_K^{(m)} = m$.
These differences between both models may provide a first idea
about the extent to which the choice of the model distribution
affects our conclusions.

\subsection{Weighting multiparticle contributions}\label{sec4a}

The normalization 
%
%
$\omega(N)$ is not a direct physical observable, but
it determines the weights with which multiparticle correlations 
contribute to the particle spectra. To see this, we consider
the one-particle spectrum, 
  \begin{equation}
    \overline{\cal P}_N^{1}({\bf P}) = 
    \sum_{m=1}^N v_m G_m({\bf P},{\bf P})/C_m\, .
    \label{4.7}
  \end{equation}
The $m$-th order contributions $G_m/C_m$ are normalized to
one, and the weights $v_m$ add up to unity,
  \begin{mathletters}
    \label{4.8}
  \begin{eqnarray}
     1 &=& \sum_{m=1}^N v_m \, ,
     \label{4.8a}\\
     v_m &=& {{(N-1)!}\over {(N-m)!}}\, 
             {\omega(N-m)\over \omega(N)}\, C_m\,  .
     \label{4.8b}
  \end{eqnarray}
  \end{mathletters}
The lowest order contribution $G_1/C_1$, by which
the one-particle spectrum is typically approximated, contributes
a fraction $v_1$ only, the value $(1-v_1)$ characterizes the
importance of higher order contributions. For a quantitative
analysis we now determine the dependence of the normalization
$\omega(N)$ and the weights $v_m$ on the event multiplicity $N$
and the phase space density of the emission region.

We consider the
terms $C_m$, the building blocks of $\omega(N)$. For the 
present model, these are given in (\ref{4.3a}).  
The factor $(h_1^{(m)} h_2^{(m)})$ in this equation ranges between
$1$ and $2^{\footnotesize{2(m-1)}}$, and can be written as
  \begin{mathletters}
    \label{4.9}
  \begin{eqnarray}
    (h_1^{(m)} h_2^{(m)}) &=& f_{\rm corr}^{(m)} 
    \left( 1 + \sqrt{ab}\right)^{\footnotesize{2(m-1)}}\, ,
    \label{4.9a} \\
    f_{\rm corr}^{(m)} &\in& \Big\lbrack {\sqrt{a\over b}}, 
                {\sqrt{b\over a}} \Big\rbrack\, .
    \label{4.9b}
  \end{eqnarray}
  \end{mathletters}
The correction factor $f_{\rm corr}^{(m)}$
appears only linearly in the expressions for $C_m$,
rather than as an $m$-th power, and it is of order $O(1)$ (it is
exactly $f_{\rm corr}^{(m)} = 1$ for a choice of parameters
$R$ and $\Delta$ such that $a = b$). This allows for the 
approximation 
  \begin{mathletters}
    \label{4.10}
  \begin{eqnarray}
    C_m &\simeq& {\epsilon}^{m-1}\, ,
    \label{4.10a} \\
    \epsilon(R,\Delta) &=& 
    \left(
    \lbrack1 + \textstyle{\sigma^2\Delta^2\over 2}\rbrack
    \lbrack1 + \textstyle{R^2\over 2\sigma^2}\rbrack 
    \lbrack 1 + \sqrt{ab}\rbrack^2
    \right)^{\footnotesize{-3/2}}\, ,
    \label{4.10b}
  \end{eqnarray}
  \end{mathletters}
with which the normalization $\omega(N)$
takes the simple form
  \begin{eqnarray}
    \omega(N) &=& \sum_{k=1}^N\, S_N^{(k)}\, \epsilon^{(N-k)}
    \nonumber \\
    &=& (-\epsilon)^N\, {\Gamma(\textstyle{-1\over \epsilon} + 1)
         \over \Gamma(\textstyle{-1\over \epsilon} + 1 - N)}
    \nonumber \\
    &=& \prod_{k=1}^N \left({ 1 + \epsilon (k-1) }\right)\, .
    \label{4.11}
  \end{eqnarray}
Here, the combinatorical factors $S_N^{(k)}$ denote the number of 
permutations of $N$ elements which contain exactly $k$ cycles.
They are commonly refered to as Stirling numbers of the first kind.
We have used their generating function in terms of
$\Gamma$-functions~\cite{GR}.

We can now determine the weights of multiparticle contributions 
to the one- and two-particle spectra. For the 
one-particle spectrum, we find by inserting (\ref{4.11}) into 
the ZP-algorithm
  \begin{mathletters}
    \label{4.12}
  \begin{eqnarray}
    \rho_{\rm vol} &:=& N\, \epsilon \, ,
      \label{4.12a} \\
    v_1 &=& {1\over {1 + \epsilon(N-1)}} 
       \approx {1\over {1 + \rho_{\rm vol}}} \, ,
       \label{4.12b}\\
    v_m 
       &\approx& {\rho_{\rm vol}^{m-1}
         \over {(1 + \rho_{\rm vol})^m}} \, .
       \label{4.12c}
  \end{eqnarray}
  \end{mathletters}
The approximation in the last line is valid for
large multiplicities, when $m \ll N$. Similarly, the weights 
$u_m$ for the different contributions to the
two-particle momentum spectrum can be calculated. Using the
power law behaviour $C_m = \epsilon^{m-1}$, we find
  \begin{mathletters}
    \label{4.13}
  \begin{eqnarray}
    && \overline{\cal P}_N^{2}({\bf P}_1,{\bf P}_2) =
    \sum_{m=2}^N u_m 
    \sum_{i=1}^{m-1}
    H_{i,m-i}({\bf P}_1,{\bf P}_2)\, ,
    \label{4.13a} \\
    &&H_{i,m-i}({\bf P_1},{\bf P_2}) = 
         {G_i({\bf P}_1,{\bf P}_1)\over C_i}
         {G_{m-i}({\bf P}_2,{\bf P}_2)\over C_{m-i}}
    \nonumber \\
    && \qquad \qquad \, 
    + {G_i({\bf P}_1,{\bf P}_2)\over C_i}
      {G_{m-i}({\bf P}_2,{\bf P}_1)\over C_{m-i}} \, ,
    \label{4.13b}
  \end{eqnarray}
  \end{mathletters}
where
  \begin{eqnarray}
    u_m &=& { (N-2)!\over (N-m)!}\,
            { \omega(N-m)\over \omega(N)}\,
            \epsilon^{m-2}
            \nonumber \\
        &\approx&
        { \rho_{\rm vol}^{m-2} \over 
          (1 + \rho_{\rm vol})^m}
        = { v_{m-1} \over 
          (1 + \rho_{\rm vol})}\, .
        \label{4.14}
  \end{eqnarray}
Again, the approximation in the last line is valid for
large multiplicities, when $m \ll N$. To sum up: 
multiparticle correlations account for a fraction
$\rho_{\rm vol} / (1 + \rho_{\rm vol})$ of the 
one-particle spectrum. For the two-particle
spectrum, they are somewhat more important:
the pure pairwise correlations receive
only a weight $1/ (1 + \rho_{\rm vol})^2$.

For sufficiently large event multiplicities, the weights $v_m$ and
$u_m$ of multiparticle contributions are not separate functions of
$\epsilon$ and $N$, but functions of the product $\rho_{\rm vol}$
only. The physics entering $\rho_{\rm vol}$ can be most easily
illustrated in the large phase space volume limit, when
  \begin{equation}
    \epsilon \approx {1\over (R^3\Delta^3)}\, ,\qquad
    \hbox{for}\, \, \, {R\over \sigma}\, , {\Delta\cdot\sigma} \gg 1\, .
    \label{4.15}
  \end{equation}
We hence call $\rho_{\rm vol}$ a ``phase space density of emission 
points''. This notion should not be taken too literally:
the product of the volumes of three-dimensional spheres in position
and momentum space is $(\textstyle{4\over 3}\pi)^2R^3\Delta^3$, and 
hence, $\rho_{\rm vol}$ is for large sources approximately a factor
10 larger than the particle number per unit phase space cell. Also, for
realistic source sizes, the value of $\epsilon$ deviates significantly from
the approximation (\ref{4.15}), and a calculation of $\rho_{\rm vol}$
starting from (\ref{4.10b}) is preferable. 

One can ask whether in the large $N$ limit, the normalization 
$\omega(N)$ becomes a function of $\rho_{\rm vol}$ only. To clarify 
this, we recall that a product 
$\prod_{k=1}^N \left( 1 + a_k\right)$ with $a_k \geq 0$ has
a $N\to\infty$-limit if and only if $\sum_{k=1}^{\infty} a_k$
converges. For the normalization (\ref{4.11}), we find
$\sum_{k=1}^N a_k$ 
$= \textstyle{1\over 2}\, \epsilon\, N\, (N-1)$, i.e.,  
for fixed phase space density,
  \begin{equation}
    \lim_{N\to \infty} \omega(N) \vert_{\rho_{\rm vol} =\, \,  
      {\rm fixed}} \longrightarrow \infty\, .
    \label{4.16}
  \end{equation}
There is no physical reason why $\omega(N)$ should remain finite.
It is not an observable. What matters in a quantitative study of
multiparticle correlation effects is the phase space density
of emission points and not the particle multiplicity, what
matters are the weights $v_m$ and $u_m$, and not the 
normalization $\omega(N)$.

We finally estimate realistic values for the phase space
density $\rho_{\rm vol}$ in heavy ion collisions.
For a choice of model parameters 
$R \approx 5$ fm, $\sigma \approx 1$ fm, $\Delta \approx 150$ MeV say,
we find $\epsilon \approx 10^{-2}$. With multiplicities of 
like-sign pions in the hundreds, this leads to $\rho_{\rm vol} > 1$. 
The present static, spherically symmetric model is however unrealistic
in so far that it does not take the strong longitudinal
expansion into account which significantly increases the
volume out of which particles are emitted. From these heuristic
considerations, we expect realistic phase space densities to
lie in the range
  \begin{equation}
    0.1 < \rho_{\rm vol} = \epsilon N < 1.0\, .
    \label{4.17}
  \end{equation}
Depending on the precise value of $\rho_{\rm vol}$ in this
range, the importance of multiparticle contributions to 
the one- and two-particle spectrum varies significantly.
For $\rho_{\rm vol} = 1.0$, higher order contributions
start dominating, while they account for $\approx$ 10 \% of the signal 
if $\rho_{\rm vol} = 0.1$.

\subsection{The momentum dependence of multiparticle contributions
}\label{sec4b}

\begin{figure}\epsfxsize=8cm 
\centerline{\epsfbox{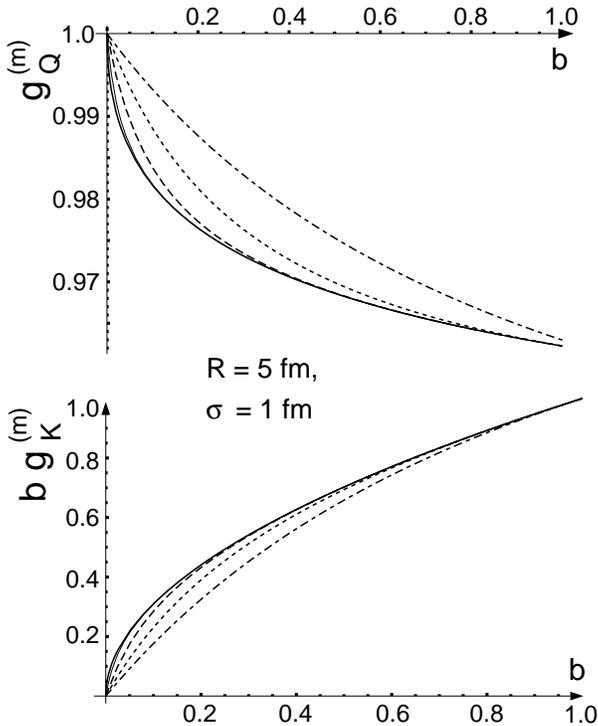}}
\caption{\it Numerical calculation of the factors $g_Q^{(m)}$ 
and $b\, g_K^{(m)}$ which determine the ${\bf q}$- and 
${\bf K}$-dependence of the $m$-th order Pratt terms. For 
a source with spatial radius $R = 5$ fm and a wavepacket 
width $\sigma = 1$ fm, the plot shows $g_Q^{(m)}$ and 
$b\, g_K^{(m)}$ as a function
of $b = 1/(1 + 2/\sigma^2\Delta^2)$. Different lines denote
different orders, $m = 2$ (dash-dotted), $m = 3$ (dotted), 
$m =5$ (dashed), $m = 10$ (thin solid) and $m = 100$ (solid). 
}\label{fig2}
\end{figure}

The two $m$-dependent terms $g_K^{(m)}$ and $g_Q^{(m)}$ in (\ref{4.4})
control the dependence of $G_m({\bf P}_1,{\bf P}_2)$ on the relative
pair momentum ${\bf q}$ and the average pair momentum ${\bf K}$.
They determine the momentum dependence of all particle spectra.
In Figure~\ref{fig2}, these factors are shown as 
a function of $b$ for a source with $R = 5$ fm and $\sigma = 1$ fm.
The first message of this plot is that even for very high
multiparticle contributions (e.g. $m = 100$), the momentum
dependence of all building blocks $G_m$ can be calculated
exactly. Secondly, the factors $g_Q^{(m)}$ and $b\, g_K^{(m)}$
show the interesting property that irrespective of the value 
chosen for $\Delta$ (and hence for $b$), they
rapidly converge to an $m$-independent quantity.
In contrast to the limiting case (\ref{4.5a}), the $m$-dependence 
of $g_Q^{(m)}$ is much weaker for realistic
model parameters $a$, $b$,  
  \begin{equation}
    g_Q^{(m)} \longrightarrow g_Q\, .
    \label{4.18}
  \end{equation}
Analogously, for realistic model parameters $a$, $b$, 
  \begin{equation}
    b\, g_K^{(m)} \longrightarrow b\, g_K\, ,
    \label{4.19}
  \end{equation}
while for the limit (\ref{4.5b}) of the parameter space, 
a strong $m$-dependence remains.
%
%
We have checked that the conclusions (\ref{4.18}), (\ref{4.19}) 
hold for a large range of the model parameter space, including 
the part realistic for heavy ion collisions. 
Going to smaller source sizes $R$ (and hence to smaller values
for $a$), the factor $g_Q$ is found to deviate significantly from
unity. Also, values for $b\, g_K$ vary significantly. The rapid 
convergence to the limiting behaviour (\ref{4.18}), (\ref{4.19}) 
however is observed for all values $a > 0.1$.
For choices of the model parameters realistic for heavy ion 
collisions, the factors $b\, g_K^{(m)}$ and $g_Q^{(m)}$ are hence
well approximated by an $m$-independent constant for sufficiently
large $m$. We can write the higher order Pratt terms as
  \begin{mathletters}
    \label{4.20}
  \begin{eqnarray}
    G_m({\bf P}_1,{\bf P}_2) &=&
    C_m\, f_g\, \exp \lbrack -A{\bf q}^2 - B{\bf K}^2 \rbrack \, ,
    \label{4.20a}    \\
    f_g &=& 
    \left( {2\pi\over \Delta^2} b\, g_K\right)^{\footnotesize{3/2}}
    \label{4.20b} \\
    A &=& \left( {\sigma^2\over 4} + {R^2\over 8} \right)\, g_Q\, ,
    \label{4.20c}    \\
    B &=& {2b\over \Delta^2}\, g_K\, .
    \label{4.20d}   
  \end{eqnarray}
  \end{mathletters}
In subsection~\ref{sec4d}, we exploit this simple $m$-dependence
of the terms $g_K^{(m)}$, $g_Q^{(m)}$.

\subsection{The one-particle spectrum 
}\label{sec4c}

For the discussion of the one-particle spectrum, we  
introduce the temperature $T$ via 
  \begin{equation}
    \Delta^2 = 2MT\, .
    \label{4.21}
  \end{equation}
The model $\rho({\bf p},{\bf r})$ in (\ref{4.1}) describes then 
a phase space distribution of emission points with Boltzmann 
temperature $T$. Our aim is to determine how the slope and shape of the 
observed one-particle spectrum $\overline{\cal P}_N^1({\bf P})$
changes with the slope $T$ of this distribution  $\rho({\bf p},{\bf r})$
and to what extent it is affected by multiparticle correlations.
$\overline{\cal P}_N^1({\bf P})$ is 
a superposition of Gaussians of different widths
  \begin{mathletters}
    \label{4.22}
  \begin{eqnarray}
    \overline{\cal P}_N^1({\bf P}) &\propto&
    \sum_{m=1}^N v_m\, {\rm e}^{-E_{\bf P}/T_{\rm eff}(m)}\, ,
    \label{4.22a} \\
    T_{\rm eff}(1) &=&
    T_{\rm eff}^{\rm pair}
    = T + {1\over 2M\sigma^2}\, ,
    \label{4.22b} \\
    T_{\rm eff}(m)
    &=& {T\over 2\, b\, g_K^{(m)}} \, \qquad m > 1\, ,
    \label{4.22c}
  \end{eqnarray}
  \end{mathletters}
where $E_{\bf P} = {\bf P}^2/2M$, and $T_{\rm eff}(m)$ characterizes 
the slope of the $m$-th order contribution. 
According to (\ref{4.22}), the
one-particle spectrum cannot be characterized by a single slope 
parameter. For a qualitative understanding, we consider first
the largest slope parameter $T_{\rm eff}^{\rm pair} = T_{\rm eff}(1)$
and the smallest slope parameter $T_{\rm eff}^{\rm mult} = 
T_{\rm eff}(m\gg 1)$. Here, the superscripts {\it pair} and {\it mult}
stand for `pairwise' and `multiparticle' correlations.

\begin{figure}\epsfxsize=8cm 
\centerline{\epsfbox{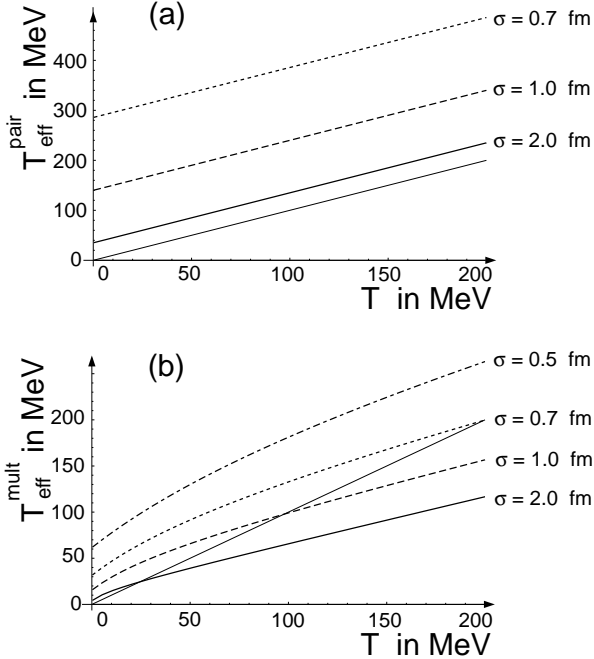}}
\caption{\it The one-particle slope parameters $T_{\rm eff}^{\rm pair}$
and $T_{\rm eff}^{\rm mult}$ characterize the limiting cases of vanishing
and dominant multiparticle correlation effects. They are shown as
functions of the model temperature $T$ for different values of 
the wavepacket width $\sigma$. The diagonals for $T_{\rm eff} = T$ (thin
solid lines) are included to guide the eye.
}\label{fig3}
\end{figure}

If all multiparticle contributions vanish, then the momentum
distribution of $\overline{\cal P}_N^1({\bf P})$ coincides with that
of $G_1({\bf P},{\bf P})$. The slope $T_{\rm eff}^{\rm pair}$
of $G_1({\bf P},{\bf P})$ is shown in Figure~\ref{fig3}a as 
a function of the model temperature $T$ for the pion mass 
$M = 139$ MeV and different
wavepacket localizations $\sigma$. $T_{\rm eff}^{\rm pair}$ is always 
larger than the model temperature $T$. For a
spatial wavepacket width $\sigma = 1$ fm e.g., the term 
${1\over 2M\sigma^2}$ takes a value of $140$ MeV. Even for 
small model temperatures $T$ as input, the
quantum contribution ${1\over 2M\sigma^2}$ accounts 
for a slope parameter $T_{\rm eff}^{\rm pair}$ comparable to the Hagedorn
temperature. For this apparently leading effect, the notion
`quantum temperature' was coined \cite{MP97}.
%
%
Figure~\ref{fig3}b shows that multiparticle contributions can
change this picture qualitatively, if they are dominant. The slope 
parameter $T_{\rm eff}^{\rm mult}$ still depends significantly on the choice 
of the wavepacket localization. But for model temperatures in the
range $100$ MeV $< T <$ $200$ MeV, there is always a value
for $\sigma$, $0.7$ fm $< \sigma <$ $1$ fm, such that the 
observed temperature $T_{\rm eff}^{\rm mult}$ coincides with
the model temperature $T$. For sufficiently large $\sigma$, the
multiparticle effect can even overcompensate the broadening due
to the quantum mechanical localization, $T_{\rm eff}^{\rm mult} < T$.
This illustrates that multiparticle
symmetrization effects tend to populate the low momentum region
of the one-particle spectrum, thereby increasing the slope of the 
spectrum. Both effects, this narrowing and the broadening due to
the quantum mechanical localization, are governed by the 
same scale $\sigma$ and hence they cancel at least to some
extent.

\begin{figure}\epsfxsize=8cm 
\centerline{\epsfbox{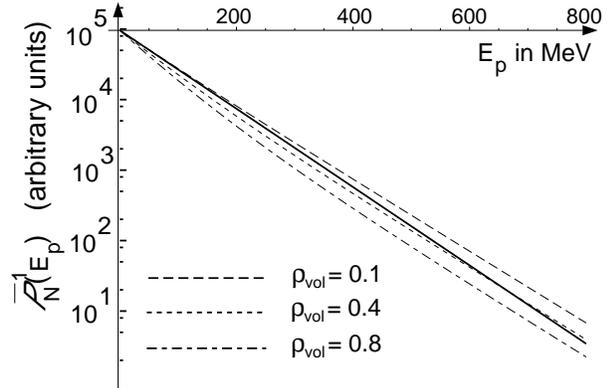}}
\caption{\it The one-particle spectrum of a source  typical
for heavy ion collisions ($R = 5$ fm, $T = 100$ MeV, $\sigma = 1.2$ fm)
becomes steeper with increasing phase space density $\rho_{\rm vol}$
of emission points. The solid line characterizes a monoexponential
behaviour and is included for comparison. 
}\label{fig4}
\end{figure}

The lowest order term $G_1$ with slope parameter
$T_{\rm eff}^{\rm pair}$ contributes a fraction $1/(1+\rho_{\rm vol})$
to the one-particle spectrum only. Hence, the one-particle spectrum 
$\overline{\cal P}_N^1$ is not monoexponential, but can be
be characterized by local slope parameters which specify
the tangent onto $\overline{\cal P}_N^1$ at $E_{\bf P} = {\bf P}^2/2M$.
Going to very large 
values of $E_{\bf P}$, the local slope always coincides with that 
of $T_{\rm eff}^{\rm pair}$, since the exponential of slowest decrease 
dominates in (\ref{4.22a}). For sufficiently small $E_{\bf P}$ 
(below $2$ GeV say), however, and for realistic phase space densities, 
neither the pair nor the multiparticle contributions can be neglected. 
The local slope lies between $T_{\rm eff}^{\rm pair}$ and 
$T_{\rm eff}^{\rm mult}$. For a more quantitative statement, we have
plotted in Figure~\ref{fig4} the one-particle spectrum
for different phase space densities $\rho_{\rm vol}$ as a function
of $E_{\bf P}$. Increasing  $\rho_{\rm vol}$ and
hence the contribution of multiparticle correlations, the local slope
of the spectrum becomes steeper. Also, the superposition of Gaussians 
of different width leads to a slight curvature of the spectrum. 
Fitting the spectra in the range 0 - 1 GeV naively with a 
monoexponential thermal distribution, a variation of $\rho_{\rm vol}$
between $0$ and $0.5$ results in a change
of the slope parameter
  \begin{equation}
    \Delta T_{\rm eff} \approx 20\, \hbox{MeV.}
    \label{4.23}
  \end{equation}
This has to be taken properly into account in a quantitative analysis
of hadron spectra.
%
%
We finally compare the spectrum (\ref{4.22a}) to that of a
Bose-Einstein distribution as obtained from an arbitrary initial
phase space distribution $\rho({\bf p},{\bf r})$, keeping the 
bosons in a box till they have equilibrated,
  \begin{mathletters}
    \label{4.24}
  \begin{eqnarray}
    {1\over {{\rm e}^{({\bf P}^2/2M - \mu)/T} - 1} }
    &=& \sum_{m=1}^{\infty}\, v_{\rm BE}^m
    {\rm e}^{-m({\bf P}^2/2M - \mu)/T}\, , 
    \label{4.24a} \\
    v_{\rm BE} &=& {\rm e}^{\mu/T}\, .
    \label{4.24b}
  \end{eqnarray}
  \end{mathletters}
For the spectrum $\overline{\cal P}_N^1$ in (\ref{4.22})
being of Bose-Einstein form, the weights $v_m$ and the 
Pratt terms $G_m$ would have to show a particular $m$-dependence,
  \begin{mathletters}
    \label{4.25}
  \begin{eqnarray}
    {\rm const}\cdot v_m &=& {v_{\rm BE}}^m\, ,
    \label{4.25a} \\
    G_m({\bf P}, {\bf P})/C_m &=& \exp \left[ -m\, E_{\bf P}/T\right]\, .
    \label{4.25b}
  \end{eqnarray}
  \end{mathletters}
In the present model, (\ref{4.25a}) is satisfied by setting
$v_{\rm BE}$ $= \rho_{\rm vol}/(1+ \rho_{\rm vol})$, see (\ref{4.12b}).
However, the $m$-dependence of the terms $G_m({\bf P}, {\bf P})$ is in
general weaker than what is required to match (\ref{4.25b}).
For the Zajc model, the $m$-dependence of $G_m$ 
is compatible with the Bose-Einstein distribution in the infinite
volume limit. However,
the Pratt terms $C_m$ differ significantly from a power law
which indicates a deviation from (\ref{4.25a}). 
The reason for these differences is, that our model calculations of
$\overline{\cal P}_N^1$ take a particular distribution 
$\rho({\bf p},{\bf r})$ as initial condition and include 
Bose-Einstein symmetrization, but they do not contain an 
equilibration mechanism: the particles are emitted and propagate
freely. Also, the $N$-particle state $\Psi_N$ is not
an equilibrium state.
Hence, the spectrum $\overline{\cal P}_N^1$ depends in 
contrast to (\ref{4.24}) on $\rho({\bf p},{\bf r})$. In general, it 
is not a Bose-Einstein distribution. 
%
\subsection{The two-particle correlator}\label{sec4d}

In this section, we first discuss the normalization of the 
two-particle correlator $C({\bf K},{\bf q})$ and how it is 
calculated in the so-called pair approximation. Then we turn
to the study of multiparticle effects.

\subsubsection{Normalization of the two-particle correlator}\label{sec4d1}

There has been some debate recently about the appropriate 
normalization of the two-particle correlation function used
in the HBT analysis of multiparticle production. For a 
compilation of the different normalizations used, and their
problems, see \cite{MV97}. In the present work, we use the
normalization~\cite{GKW79}
  \begin{equation}
    C({\bf P}_1,{\bf P}_2) =
    { {\langle \hat{N}\rangle^2}\over 
      {\langle \hat{N}(\hat{N} - 1) \rangle}}
    { \sigma_{\pi}\, \, d^6\sigma_{\pi\pi}/ d^3{\bf P}_1\,d^3{\bf P}_2 
      \over (d^3\sigma_{\pi}/ d^3{\bf P}_1)\, 
            (d^3\sigma_{\pi}/ d^3{\bf P}_2) }\, ,
    \label{4.26}
  \end{equation}
where $\sigma_{\pi}$ is the total pion cross section.
This originates from normalizing both the single- and the
double-differential cross sections to unity. Setting these
cross sections proportional to the one- and two-particle
spectra $\overline{\cal P}_N^1({\bf P}_1)$, 
$\overline{\cal P}_N^2({\bf P}_1,{\bf P}_2)$, 
%
%
the two-particle correlator reads
  \begin{equation}
    C({\bf P}_1,{\bf P}_2) =
    { \overline{\cal P}_N^2({\bf P}_1,{\bf P}_2)
      \over \overline{\cal P}_N^1({\bf P}_1)
            \overline{\cal P}_N^1({\bf P}_2) }\, .
    \label{4.28}
  \end{equation}
The simplest way to make further progress is to assume that the
approximation (\ref{4.20}) for $G_m$ is valid for all $m \geq 1$.
Then, the two-particle spectrum can be written as
  \begin{eqnarray}
    &&\overline{\cal P}_N^2({\bf P}_1,{\bf P}_2) =
    {\tilde{\omega}(N)\over \omega(N)}\, f_g\, 
    {\rm e}^{-A\, {\bf q}^2 -B\, {\bf K}^2}\, ,
    \nonumber \\
    &&\tilde{\omega}(N) = \sum_{J=2}^N {(N-2)!\over (N-J)!}\, 
    \omega(N-J)\, \sum_{i=1}^{J-1}\, C_i\, C_{J-i}\, ,
    \label{4.29}
  \end{eqnarray}
and the correlator reads
  \begin{equation}
    C({\bf P}_1,{\bf P}_2) =
    {\tilde{\omega}(N)\over \omega(N)}
    \left( 1 + { {\rm e}^{-2A{\bf q}^2 - 2B{\bf K}^2}
               \over {\rm e}^{-B{\bf P}_1^2}\, {\rm e}^{-B{\bf P}_2^2}}
             \right)\, .
    \label{4.30}
  \end{equation}
The normalization ${\tilde{\omega}(N)\over \omega(N)}$ 
obtained in this approximation 
remains unchanged if the full $m$-dependence of the
Pratt terms $G_m$ is included, though the momentum dependence of
(\ref{4.30}) is then much more involved.

The structure of the normalization ${\tilde{\omega}(N)\over \omega(N)}$
is important for what follows: The term $\tilde{\omega}(N)$
contains exactly $N!/2$ terms, while $\omega(N)$ contains $N!$ terms.
By integrating $\omega(N) \bar{\cal P}_N^2({\bf P}_1,{\bf P}_2)$
over ${\bf P}_2$ and using the power law
behaviour $C_m = \epsilon^{m-1}$, we find
  \begin{mathletters}
    \label{4.31}
  \begin{eqnarray}
    \omega(N) - \tilde{\omega}(N)
    &=& \sum_{J=2}^N {(N-2)!\over (N-J)!}\, 
    \omega(N-J)\, \sum_{i=1}^{J-1}\, C_J\, ,
    \label{4.31a} \\
    &&{\tilde{\omega}(N)\over \omega(N)} = 1 - \epsilon\, .
    \label{4.31b}
  \end{eqnarray}
  \end{mathletters}
The normalization of the correlator is smaller than unity, the
offset depends on the phase space volume occupied by the source.
We shortly comment on the importance of this result:

The two-particle correlator can be viewed as the factor relating
two-particle differential cross sections $\sigma_{\rm BE}$ of 
the real world (where Bose-Einstein symmetrization exists) to
an idealized world in which Bose-Einstein correlations are
absent,
  \begin{equation}
    \sigma_{\rm BE}({\bf K},{\bf q}) = C({\bf K},{\bf q})\, 
    \sigma_{\rm NO}({\bf K},{\bf q})\, .
    \label{4.32}
  \end{equation}
If $C({\bf K},{\bf q}) > 1$ everywhere, this implies that
Bose-Einstein symmetrization effects increase the total
cross section. For heavy ion collisions, there is no direct
test of whether such an enhancement exists. In $e^+e^-$-collisions
however, we know that Bose-Einstein correlations do not affect
total cross sections appreciably, since e.g. perturbative
QCD predicts the production characteristics of $Z_0$ to a
per mille level without invoking them. In our calculation,
we start from an $N$-particle state and we require that after
final state Bose-Einstein symmetrization, $N$ particles are
detected. Hence, we explicitly assume that final state
Bose-Einstein correlations do not affect the total cross
sections. According to the above calculation, this automatically
implies an offset of the normalization of $C({\bf K},{\bf q})$
below unity. The offset $\epsilon$ in (\ref{4.31b})
measures the system size from which particles are
emitted. This is intrinsically consistent: for smaller system
sizes ($\epsilon$ large), the correlator shows an enhancement
in a broader ${\bf q}$ momentum region, and the normalization
$(1-\epsilon)$ is smaller. This ensures that $\sigma_{\rm BE}^{\rm tot}
= \sigma_{\rm NO}^{\rm tot}$. 

We finally note that irrespective of the offset $\tilde{\omega}(N)/
\omega(N)$ of the normalization, the correlator in (\ref{4.30})
changes by a factor $2$ between the limits ${\bf q} = 0$ and
${\bf q} \to \infty$. This is a consequence of the approximation
(\ref{4.20}) which we have justified for the present model in
subsection~\ref{sec4b}. In general, the ${\bf K}$- and ${\bf q}$-
dependence of the Pratt terms $G_m$ leads to a more complicated
dependence of the correlator $C({\bf K},{\bf q})$ according to
(\ref{4.13}). Depending on the model, this may affect the intercept
of $C({\bf K},{\bf q})$ at ${\bf q} = 0$. Indeed, a decrease of 
the intercept with increasing event multiplicities  was reported
in recent model studies~\cite{P93,CZ97}. In contrast to the present 
study, these models however do not work with fixed event multiplicities and
show a significantly different physics, including pion lasing
effects~\cite{P93}. We only conclude from the present study that 
multiparticle symmetrization effect do not lead automatically to 
a strong decrease of the intercept parameter.

\subsubsection{The normalization in the pair approximation}\label{sec4d2}

We now explain why the offset $(1-\epsilon)$ of the normalization of
$C({\bf K},{\bf q})$ is not obtained in conventional calculations where
multiparticle symmetrization effects are neglected.
For the wavepackets introduced in section~\ref{sec3}, the
wavepacket overlap functions $f_{ij}$ in (\ref{3.4}) shows a 
Gaussian decrease with the phase space distance $\vert z_i - z_j\vert$.
In general, the
  \begin{equation}
    \hbox{pair approximation:  }\, f_{ij} = \delta_{ij}
    \label{4.33}
  \end{equation} 
is an uncontrolled approximation to this Gaussian behaviour. 
It becomes exact in certain limiting cases
  \begin{equation}
    \lim_{\sigma \to 0} f_{ij} = \lim_{\sigma \to \infty} f_{ij}
    = \delta_{ij}\, .
    \label{4.34}
  \end{equation}
In the pair approximation, all higher order Pratt terms vanish
  \begin{eqnarray}
    &&C_m = G_m({\bf K},{\bf q}) = 0\, , \qquad m \geq 2\, ,
    \nonumber \\
    && \hbox{for}\qquad f_{ij} = \delta_{ij}\, ,
    \label{4.35}
  \end{eqnarray}
and one regains the results of the conventional calculations where
multiparticle effects are neglected: the two-particle correlator is 
normalized to unity,
  \begin{equation}
    \omega(N) = \tilde{\omega}(N) = 1\, \qquad 
    \hbox{for}\, \,  f_{ij}=\delta_{ij}\, ,
    \label{4.36}
  \end{equation}
and (\ref{2.5ext}) does not change the phase space
distribution since $\omega(N) = \langle\Psi_N\vert\Psi_N\rangle$. The term
$G_1({\bf P}_1,{\bf P}_2)\, G_1({\bf P}_1,{\bf P}_2)$ in (\ref{4.13})
then insures that $C({\bf P}_1,{\bf P}_2) > 1 $. According to
(\ref{4.32}), this contradicts the statement
$\sigma_{\rm BE}^{\rm tot} = \sigma_{\rm NO}^{\rm tot}$.
The origin of this difficulty can be traced back to the integral
  \begin{equation}
   \int d^3{\bf P}_1\, d^3{\bf P}_2\,
   G_1({\bf P}_1,{\bf P}_2)^2 = C_2\, ,
   \label{4.37}
  \end{equation}
which should vanish according to (\ref{4.35}). The pair approximation
does not treat the integral $C_2$ and the integrand of (\ref{4.37})
on an equal footing. It sets $C_2 = 0$ but uses 
$G_1({\bf P}_1,{\bf P}_2)^2$ for the calculation of the Bose-Einstein
enhancement. It is this inconsistency which leads to a correlator
$C({\bf K},{\bf q}) > 1$ everywhere.

\subsubsection{The HBT-radius parameters}\label{sec4d3}

Once the momentum-dependent higher order Pratt terms $G_m$
and their weights $v_m$, $u_m$ are known, the determination
of the two-particle correlator (\ref{4.28}) is a matter of
straightforward numerical calculation. From $C({\bf K},{\bf q})$,
the HBT radius parameters are obtained by fitting
the Gaussian ansatz
  \begin{equation}
    C({\bf K},{\bf q}) \propto 1 + \lambda 
    \exp\left[-R_{\rm HBT}^2{\bf q}^2\right]\, .
    \label{4.38}
  \end{equation}
We restrict our discussion to this one-dimensional
parametrization of $C({\bf K},{\bf q})$ since our model (\ref{4.1})
is spherically symmetric. The general, analytical form of the
two-particle correlator (\ref{4.28}) is obtained from (\ref{4.7})
and (\ref{4.13}) and it is quite involved. For a transparent
discussion, we hence turn first to limiting cases. 
In the pair approximation, when all multiparticle effects are
neglected, we obtain from (\ref{4.2b}) the well-known 
result~\cite{Weal97}
  \begin{equation}
    {R_{\rm HBT}^{\rm pair}}^2
    = {R^2\over 2} + {\sigma^2\over 2}
    {{\Delta^2\sigma^2}\over {1 + {\Delta^2\sigma^2}}}\, .
    \label{4.39}
  \end{equation}
The other, equally unrealistic limiting case is that 
multiparticle contributions dominate completely. 
The resulting two-particle correlator is given in 
(\ref{4.30}) and the corresponding HBT-radius parameter 
reads
  \begin{equation}
    {R_{\rm HBT}^{\rm mult}}^2 = {R^2\over 4}g_Q +
    {\sigma^2\over 2} 
    {{g_Q\sigma^2\Delta^2 + 2g_Q - g_K}\over 
      {\sigma^2\Delta^2 + 2}}\, ,
    \label{4.40}
  \end{equation}
The difference between these two expressions is significant. 
For a discussion of the main qualitative and quantitative
effects, we now focus on the parameter region $R^2 \gg \sigma^2$
relevant for relativistic heavy ion collisions. In this regime,
the $\textstyle{\sigma^2\over 2}$-dependent parts of (\ref{4.39})
and (\ref{4.40}) can be neglected, and according to the study
of section~\ref{sec4b}, $g_Q \approx 1$. We then find the
simple relation
  \begin{equation}
    R_{\rm HBT}^{\rm mult} \approx {1\over \sqrt{2}}
    R_{\rm HBT}^{\rm pair}\, ,\qquad \hbox{for}\qquad
    R \gg \sigma\, .
    \label{4.41}
  \end{equation}
This is the factor $2$, obtained first by Zajc~\cite{Z87}
in his comparison of the first and second order Pratt terms,
see (\ref{2.16}). Zajc has given an estimate that 
in his model, multiparticle effects change the HBT radius 
parameter by as much as a factor 0.67 for a particle density
of $1$ per unit phase space cell~\cite{Z87}. This would affect estimates of
the source volume or the energy density by a factor 3.
In Zajc's model, however, higher order multiparticle
effects ($m>2$) change the ${\bf q}$-dependence of $C({\bf K}, {\bf q})$
much more dramatically than in the present study, where 
$g_Q^{(m)}$ has a very weak $m$-dependence, negligible
for $R^2 \gg \sigma^2$. 

\begin{figure}\epsfxsize=8cm 
\centerline{\epsfbox{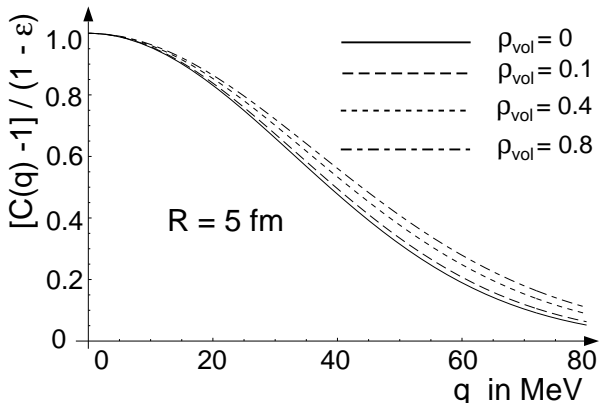}}
\caption{\it Multiparticle symmetrization effects broaden the
two-particle correlator. For a source size of $R = 5$ fm, the
plot shows the resulting two-particle correlator for different
particle phase space densities $\rho_{\rm vol}$.
}\label{fig5}
\end{figure}

In our model, the correlator is a weighted superposition of 
Gaussians in ${\bf q}$
whose widths differ by as much as a factor $\sqrt{2}$.
In Figure~\ref{fig5}, we have plotted the resulting two-particle
correlator for relatively low phase space densities where the
lowest order Pratt term $G_1$ is still the leading contribution.
The result of the pair approximation describes the main behaviour,
but deviations due to multiparticle contributions affect the HBT-radius
parameter on a 10 \%-level for moderate phase space densities.
This 10\% effect translates into underestimating 
the volume by 30 \% and overestimating the corresponding energy 
density by a similar amount. In models which show a stronger 
$m$-dependence of the momentum-dependent part of $G_m$, the effect 
may be significantly stronger. In general, 
the degree to which multiparticle effects modify
HBT radius parameters depends significantly on the phase space
density of emission points. 
%
%

\section{Multiparticle correlations for arbitrary models}\label{sec5}

In this section, we shortly discuss how the calculation of 
multiparticle correlation effects can be extended to more 
realistic source distributions where the analytical techniques 
used in section~\ref{sec4} are not applicable. Pratt has shown
already how to calculate the terms $C_m$ and $G_m({\bf P}_1,{\bf P}_2)$
for {\it continuous} source functions~\cite{P93}: up to order $m \leq 5$,
straightforward Monte Carlo methods can be used, and
an improved Metropolis method allows to push numerical
calculations up to $m \approx 20$. This seems to be sufficient 
for all practical purposes, since according to our
model study, the weights $v_m$, $u_m$ for $m > 20$ are
negligible for realistic phase space densities
$\rho_{\rm vol}$.

Here, we propose yet a different technique which is 
applicable to events characterized by a set of $N$
{\it discrete} phase space points $({\bf p}_i,{\bf r}_i,t_i)$. 
We interprete these points as the centers of Gaussian 
single-particle wavepackets, i.e.,
  \begin{equation}
    ({\bf p}_i,{\bf r}_i,t_i) \longrightarrow f_i\, .
    \label{5.1}
  \end{equation}
An arbitrary distribution of $N$ discrete phase space
points is the most general `source model' in
the present framework. We first discuss algorithms which allow
for the calculation of the two-particle correlator from a set
$({\bf p}_i, {\bf r}_i, t_i)$. Then we comment on applications
of these algorithms to event generators.

\subsection{An algorithm for Bose-Einstein weights}\label{sec5a}

We start with the simplest case, the pair approximation, 
which reduces the sum over $N!$ terms in the
two-particle spectrum to a sum over all 
$\textstyle{1\over 2}N(N-1)$ particle pairs $(i,j)$.
Each pair is weighted with the pair probability ${\cal P}_{ij}$,
calculated from the corresponding two-particle symmetrized 
state,~\cite{Weal97}
  \begin{eqnarray}
    {\cal P}_{ij}({\bf P}_1,{\bf P}_2) &=& {1\over 2}
    \Bigg\vert D_i({\bf P}_1)\, D_j({\bf P}_2)
               + D_i({\bf P}_2)\, D_j({\bf P}_1) \Bigg\vert^2\, ,
               \nonumber \\
    {\cal P}^2_N({\bf P}_1,{\bf P}_2) &=& 
    {2\over N(N-1)} \sum_{(i,j)}
    {\cal P}_{ij}({\bf P}_1,{\bf P}_2)\, .
    \label{5.2}
  \end{eqnarray}
To compare this expression to the ZP-algorithm, we introduce the
quantities
  \begin{mathletters}
    \label{5.3}
  \begin{eqnarray}
    G_1({\bf P}_1,{\bf P}_2) &=& \textstyle{1\over N}
    \sum_{i=1}^N D_i^*({\bf P}_1)D_i({\bf P}_2)\, ,
    \label{5.3a} \\
    T_c({\bf P}_1,{\bf P}_2) &=& \textstyle{1\over N^2}
    \sum_{i=1}^N D_i^*({\bf P}_1)D_i({\bf P}_1) 
    \nonumber \\
    && \qquad \times\, D_i^*({\bf P}_2)D_i({\bf P}_2)\, .
    \label{5.3b}
  \end{eqnarray}
  \end{mathletters}
Here, $G_1({\bf P}_1,{\bf P}_2)$ is the first order Pratt term 
for the discrete phase space distribution $({\bf p}_i, {\bf r}_i, t_i)$
and $T_c$ is a finite multiplicity correction, correcting for
the double counting of identical pairs $(i,i)$ in
  \begin{eqnarray}
    \sum_{(i,j)} {\cal P}_{ij}({\bf P}_1,{\bf P}_2)
    &=& G_1({\bf P}_1,{\bf P}_2)\, G_1({\bf P}_1,{\bf P}_2)
    \nonumber \\
    && + G_1({\bf P}_1,{\bf P}_2)\, G_1({\bf P}_1,{\bf P}_2)
    \nonumber \\
    && - 2T_c({\bf P}_1,{\bf P}_2)\, .
    \label{5.4}
  \end{eqnarray}
In the large multiplicity limit, when $T_c$ can be neglected,
this expression is equivalent to the pair approximation of
the ZP-algorithm, see e.g. (\ref{4.13a}). The one-particle spectrum
$\nu$ and the 2-particle correlation obtained in this way
have been discussed already extensively in~\cite{Weal97}. They
can be rewritten in terms of the single-particle probabilities 
$s_i$ of (\ref{3.3b}),
  \begin{mathletters}
    \label{5.5}
  \begin{eqnarray}
    C({\bf K},{\bf q}) &=&
    1 + {{ {\rm e}^{-\sigma^2{\bf q}^2/2}
        \Big\vert \sum_{i=1}^N s_i({\bf K})
        {\rm e}^{i{\bf r}_i\cdot{\bf q}}\Big\vert^2
        - T_c}       \over 
         {\nu({\bf P}_1)\, \nu({\bf P}_2) - T_c } }\, ,
   \nonumber \\
   \nu({\bf P}) &=&
        \sum_{i=1}^N s_i({\bf P})\, ,
        \label{5.5a} \\
    T_c({\bf P}_1,{\bf P}_2) &=& 
         \sum_{i=1}^N s_i({\bf P}_1)\, s_i({\bf P}_2)\, .
         \label{5.5b}
  \end{eqnarray}
  \end{mathletters}
The number of numerical operations needed to calculate this 
correlator grows linearly with the multiplicity
$N$ rather than quadratic as in (\ref{5.2}) and this makes
it particularly suitable for a numerical
algorithm. At given observed momenta ${\bf K}$, ${\bf q}$,
one calculates for each event the one particle probabilities
$s_i({\bf K})$ according to (\ref{3.3b}) and performes the sums
in (\ref{5.5}). The $s_i({\bf K})$ are continuous in the 
measured momentum, and hence the obtained correlator 
is a continuous function of ${\bf K}$ and ${\bf q}$, i.e., no 
binning is required.
At high multiplicities, when the factors $T_c$ are negligible, 
the correlator (\ref{5.5}) can be rewritten as a Fourier transform 
over an emission function $S(x,{\bf K})$, thereby regaining the
well-known starting point of most model studies, 
(see Refs.~\cite{Weal97,ZWSH97} for further details)
  \begin{mathletters}
  \label{5.6}
  \begin{eqnarray}
    C({\bf K},{\bf q}) &=& 1 + 
        {{\Big\vert \int d^4x\, S(x,{\bf K}) 
          {\rm e}^{ix\cdot q}\Big\vert^2}
      \over
      { \int d^4x\, S(x,{\bf P}_1)\,\,  
        \int d^4y\, S(y,{\bf P}_2)} } \, ,
   \label{5.6a} \\
    S(x,K) &=& \sum_{i=1}^\infty S_i(x,{\bf K})\, ,
    \label{5.6b}\\
    S_i(x,{\bf K}) &\propto& \delta(t-t_i)\,
    {\rm e}^{-\textstyle{1\over \sigma^2}({\bf x}-{\bf r}_i)^2
    -\textstyle{\sigma^2} ({\bf K}-{\bf p}_i)^2}\, .
    \nonumber
  \end{eqnarray}
  \end{mathletters}
The above discussion shows, that in the pair approximation,
a numerical algorithm for the calculation of one- and two-particle
spectra can be based on a discrete version of the first order
Pratt terms $G_1$. Here, we propose to extend this approach to
take multiparticle correlations into account by calculating
  \begin{mathletters}
    \label{5.7}
  \begin{eqnarray}
    &&C_m = {(N-m)!\over N!} \sum_{i_1...i_m}
            f_{i_1i_2}....f_{i_{m-1}i_m}f_{i_mi_1}\, ,
            \label{5.7a} \\
   &&G_m({\bf P}_1,{\bf P}_2) = {(N-m)!\over N!} \sum_{i_1...i_m}  
     D_{i_1}^*({\bf P}_1)\, f_{i_1i_2}..
     \nonumber \\
     && \qquad \qquad \qquad \qquad ..f_{i_{m-1}i_m}
     D_{i_m}({\bf P}_2)\, .
     \label{5.7b}
  \end{eqnarray}
  \end{mathletters}
The sums in these expressions run over all sets of $m$ out
of $N$ integers and over all $m!$ permutations of
each set. This implies that for event multiplicities in the
hundreds, only terms up to order $m \approx 5$ can be calculated 
in a reasonable amount of CPU time. A tentative strategy for 
calculating multiparticle correlation effects on the basis of
(\ref{5.7}) then proceeds as follows:
  \begin{enumerate}
    \item
      Fit the calculated terms $C_m$ to a power law 
      $C_m = \epsilon^{m-1}$ and use the parameter $\epsilon$ 
      thus determined for a calculation of the weights $v_m$, $u_m$
      in (\ref{4.7}) and (\ref{4.13a}).
   \item
     Fit Gaussians in ${\bf K}$ and ${\bf q}$ to the 
     momentum-dependent Pratt terms $G_m({\bf P}_1,{\bf P}_2)$.
     This allows to extract the terms $g_K^{(m)}$ and $g_Q^{(m)}$
     which govern the momentum dependence of multiparticle effects.
   \item
     Calculate the one- and two-particle spectra by including
     up to the numerically determined order, $m = 5$ say, all 
     contributions exactly, approximating the momentum dependence
     of higher order terms by setting $g_K^{(m)} = g_K^{(5)}$,
     $g_Q^{(m)} = g_Q^{(5)}$ for $m > 5$.
  \end{enumerate}
This scheme draws on the experience gained from the study of the
Gaussian toy model in section~\ref{sec4}. We expect that it allows 
to estimate multiparticle contributions for model distributions 
$({\bf p}_i, {\bf r}_i, t_i)$.

\subsection{Bose-Einstein weights for event generators}\label{sec5b}

Many event generators for the simulation of heavy ion collisions
have been developed in recent years~\cite{SSG89,W93,PSK92,KKG97,WG91}, 
and irrespective of the large variety of physical inputs 
present in their codes, the typical output of their
event simulation contains for each event a set 
$\Sigma$ of $N$ phase space points $({\bf p}_i,{\bf r}_i,t_i)$ 
which one associates to the final state particles produced. 
However, a choice of interpretation is involved in comparing an 
event generator output $z_i = ({\bf p}_i,{\bf r}_i,t_i)$ 
to experimental data. Usually, the measured one-particle 
spectra are compared directly to the binned momenta 
${\bf p}_i$, i.e. one implicitly interprets the simulated 
phase space points $z_i$ as defining momentum eigenstates, 
which a fortiori carry no space-time information. For the
calculation of two-particle correlations, this interpretation
is not suitable, it leads to a sharp $\delta$-like correlator
~\cite{Weal97}. Also, a classical interpretation of $z_i$ 
which takes both ${\bf r}_i$ and ${\bf p}_i$
as ``sharp'' information, is problematic:
ignoring the correct quantum mechanical localization
leads to quantitatively and qualitatively unreliable 
results~\cite{MKFW96}. On the other hand, including
a quantum mechanical localization width $\sigma$, one
changes both the one- and two-particle spectrum by a
prescription which has no dynamical foundation. 

The origin of these difficulties in finding a consistent
interpretation of the event generator output 
$z_i = ({\bf p}_i,{\bf r}_i,t_i)$ is well-known, see
e.g.~\cite{LS95,A96}: 
quantum mechanical processes require a description in terms of 
amplitudes, while event simulations are formulated in a 
probabilistic setting. Bose-Einstein correlations occur 
by symmetrizing the production amplitudes of identical 
particles and are hence not encoded for in event generators.
The mere fact that the simulated output is a discrete
phase space distribution of emission points without 
identical multiparticle correlations  (rather than a
set of detected momenta which includes all multiparticle
correlations and hence all space-time information available
from the measurement), is the very consequence of using a
probabilistic language. In practice, this forces one to
take recourse to an algorithm which associates {\it a posteriori}
Bose-Einstein weights with the simulated phase space distribution
$\Sigma$, rather than obtaining these correlations from the
dynamical propagation of properly (anti)-symmetrized N-particle 
states. This a posteriori modification of an incomplete
quantum-dynamical evolution creates the interpretational
problems of the output $z_i = ({\bf p}_i,{\bf r}_i,t_i)$.
Phenomenologically motivated numerical simulations of heavy
ion collisions have remarkable success despite
these fundamental problems. It is of some interest to confront
them with experimental two-particle correlations, since this
provides a test of their spatio-temporal (rather than only their 
momentum-dependent) properties. We expect that the 
algorithms discussed in section \ref{sec5a} are useful in such
comparisons.

\section{Conclusion}\label{sec6}

The Zajc-Pratt algorithm provides a simple technique for the
calculation of multiparticle symmetrization effects in one- and
two-particle spectra. Based on this algorithm, we have studied to what extent
multiparticle correlations steepen the slope of the one-particle
and broaden the width of the two-particle spectrum. The scale of
both effects depends sensitively on the particle phase space 
density $\rho_{\rm vol}$ in the emission region.
Multiparticle correlations contribute a fraction 
$\rho_{\rm vol}/(1 + \rho_{\rm vol})$ to the one-particle spectrum 
and a fraction $1 - 1/(1 + \rho_{\rm vol})^2$ to the two-particle 
spectrum. Also, the $m$-dependence of higher order Pratt terms, 
i.e., the extent to which the momentum dependence of multiparticle 
contributions $G_m$ changes with increasing order $m$, plays an 
important role. In a Gaussian source model with instantaneous 
emission and moderate particle phase space densities, the slope 
parameter $T_{\rm eff}$ of the one-particle spectrum changes by 
up to 20 MeV, and the change in the HBT radius parameters is of the
order of 10 percent, if multiparticle effects are
neglected. 

For the reconstruction program advocated in \cite{HW97}, our results
indicate that neglecting multiparticle effects, one 
underestimates the source temperature significantly and one
overestimates the energy density by up to 30 \%. Here, the caveat
is however that our calculation does not include multiparticle
final state Coulomb interactions. Bose-Einstein and Coulomb effects
typically arise on the same scale and compensate each other at 
least to some extent. This may significantly reduce the effect
of multiparticle correlations in the measured $\pi^+$ and 
$\pi^-$ spectra on which most of the phenomenological analysis
is based currently. It will hence be very interesting to compare
the slopes of the one-particle spectra of charged and neutral
pions~\cite{WA98}. On the basis of the above heuristic ideas, 
one may expect the slope of the $\pi^0$-spectrum to be somewhat 
steeper, since multiparticle
Coulomb interactions cannot compensate for the multiparticle
Bose-Einstein symmetrization effects. 

Our analysis of multiparticle effects has also shed new light on a
recent discussion about the normalization of the two-particle
correlator. Defining the correlator via (\ref{4.26}), it is normalized
for large relative momenta ${\bf q}$ to $(1-\epsilon)$, and not to
unity. The importance of this result was already discussed in
section~\ref{sec4d1}: the normalization
(\ref{4.31b}) corrects the result of the commonly 
used pair approximation and leads to Bose-Einstein unit weights
which conserve event probabilities.

We close by embedding the present work into a wider perspective:
Many aspects of the dynamical evolution of relativistic heavy
ion collisions are mesoscopic, and this makes it very difficult
to decide whether certain physical observables have a known
conventional interpretation or are indicative of the new in
medium properties which the current experimental programs at
CERN and RHIC aim for. The physics of final state hadrons however,
though being mesoscopic and hence difficult to treat, is known
in principle. It allows hence for a detailed quantitative
description on the basis of present day knowledge. In a second 
step, such a description will pave the way for quantitative
statements about the geometry and dynamics of the final stage
of the collision process, thereby setting up an experimentally
motivated starting point for a dynamical back extrapolation~\cite{HW97}. 
We hope that the present work, by quantifying one mesoscopic
multiparticle effect, will prove useful in refining this
reconstruction program and in separating known physics from 
the in medium properties we are looking for.

%
\acknowledgements
The author thanks Miklos Gyulassy and Bill Zajc for helpful
discussions and the hospitality extended to him at Columbia University. 
They and their groups (and especially Yang Pang and Stephen Vance) 
provided a stimulating atmosphere throughout this work. Also, it
is a pleasure to acknowledge conversations in the ``Regensburg
Cappuccino group'' with Rainer Fries and Martin Maul, concerning the
determinants of band matrices, and with Claus Slotta, concerning
the physics underlying Eq. (\ref{2.16}). Discussions with
Klaus Kinder-Geiger and Tjorbj\"orn Sj\"ostrand have sharpened 
the ideas presented in section~\ref{sec5}. Last but not least, 
the author thanks Ulrich Heinz for a critical reading of the
manuscript and a discussion focussing on Eq. (\ref{4.24}). This work 
was supported by a DFG Habilitandenstipendium and by the US 
Department of Energy under Contract No. DE-FG-02-86ER40281.

\appendix

\section{Derivation of the ZP-algorithm}
\label{appb}

So far, there is no self-contained derivation of the Zajc-Pratt
algorithm in the literature. For the convenience of the reader,
we give here the main arguments.

\begin{figure}\epsfxsize=8cm 
\centerline{\epsfbox{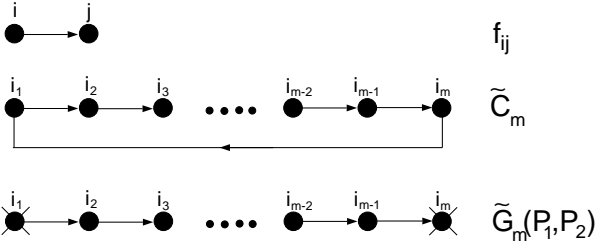}}
\caption{\it Diagrammatic representation of the types of factors
contributing to multiparticle spectra. 
}\label{fig6}
\end{figure}

\begin{figure}\epsfxsize=8cm 
\centerline{\epsfbox{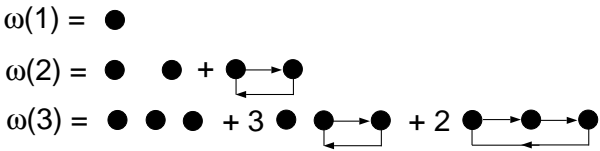}}
\caption{\it Diagrammatic representation of the averaged total
multiplicity $w(N)$. The 2-cycle stands for $C_2$, the
3-cycle for $C_3$, etc.
}\label{fig7}
\end{figure}

{\it Notation:} We denote the term $f_{i_1i_2}$ by an arrow from 
$i_1$ to $i_2$. $m$ of these terms form a closed $m$-cycle 
$\tilde{C}_m$. The only other structure 
appearing in all spectra are open $m$-cycles $\tilde{G}_m$, 
see Figure~\ref{fig6}. Here, $D_{i_1}^*$ is represented 
by a cross from which the arrow starts, $D_{i_1}$ by a cross 
at which the arrow ends. The Pratt terms $C_m$, $G_m$ are obtained 
by averaging the emission points in Figure~\ref{fig6} over a 
model-distribution $\rho({\bf p},{\bf r}, t)$. Dots without
indices attached indicate that this average was carried out,
see Figure~\ref{fig7}.
%
%
{\it Normalization:} The normalization 
$w(N) = \overline{\cal P}_N^{0}$ sums the products
$ C_1^{l_1}\, C_2^{l_2}\, ...\, C_n^{i_n}$ over all 
partitions $(n,l_n)_N$. Each partition of $(n,l_n)_N$  
has $N! / \prod_n n!^{l_n} (l_n!)$ possible realizations. 
These have to be multiplied by the $(n-1)!$ different ways to 
combine each set of $n$ points to a closed cycle. This leads to
the prefactor $N! / \prod_n n^{l_n} (l_n!)$ in (\ref{2.7}).
%
%
There exists a simple diagrammatic prescription to construct 
from the terms of $w(N)$ the terms of $w(N+1)$. One adds
to each diagram of $w(N)$ an $(N+1)$th dot by
  \begin{itemize}
    \item
      1.) placing the dot once disconnected from all other cycles.
    \item
      2.) changing closed $m$-cycles $C_m$ into $m$ closed
      $(m+1)$-cycles $C_{m+1}$.
  \end{itemize}
This prescription can be checked e.g. in Figure~\ref{fig7}. It
automatically insures, that $w(N)$ is
represented by $N!$ diagrams, the maximal value
for $w(N)$ is $N!$. 

\begin{figure}\epsfxsize=8cm 
\centerline{\epsfbox{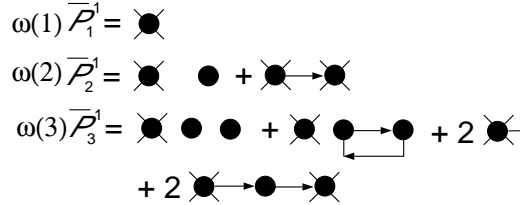}}
\caption{\it Diagrammatic representation of the 1-particle spectrum
for multiplicities $N = 1$,
$2$, $3$.
}\label{fig8}
\end{figure}

{\it One-particle spectrum}:  Each term of
$\omega(N)\, \overline{\cal P}_N^{1}({\bf P})$ has the structure
$w(N-m)\, G_m$: it contains exactly one open
$m$-cycle $G_m(P,P)$, the $N-m$ other dots being contained in
closed cycles. To construct from it the terms of 
$\omega(N+1)\, \overline{\cal P}_{N+1}^{1}({\bf P})$, one has to
use the replacement
  \begin{eqnarray}
    w(N-m)\, G_m \longrightarrow && 
      w(N+1-m)\, G_m
      \nonumber \\ 
      &&+ w(N-m)\, m\, G_{m+1}\, .
      \label{B1}
  \end{eqnarray}
Diagrammatically, this is realized by employing the two diagrammatic
rules given above and supplementing them with
  \begin{itemize}
    \item
      3.) change open $m$-cycles $G_m$ into $m$ open
      $(m+1)$-cycles $G_{m+1}$.
  \end{itemize}
The prescription can be checked e.g. in
Figure~\ref{fig8}. Using the recursion relation (\ref{B1}), one can 
prove by complete induction that the averaged one-particle spectrum
$\overline{\cal P}_N^{1}({\bf P})$ takes the form (\ref{2.8}).
%
%
{\it Two-particle spectrum:} In 
$\omega(N)\, \overline{\cal P}_N^{2}({\bf P_1},{\bf P_2})$, each term contains
two open cycles of lengths $i$ and $j$, the $N-i-j$ remaining dots
being contained in closed cycles. On the endpoints of the two 
open cycles $G_i$ and $G_j$, the momenta ${\bf P}_1$ and ${\bf P}_2$ 
can be attached in different combinations: for $i=j$, there are two
possiblities, for $i\not= j$, there are four. This combinatorics
is properly encoded in the definition of the product $G_{i,j}$
$ = C_i C_j H_{i,j}$ $/(1+\delta_{ij})$, see Eq. (\ref{4.13b}).
An $N \to (N+1)$ recursion relation for the two-particle
spectrum is obtained by using the rule to change 
open $m$-cycles into $m$ open $(m+1)$-cycles, and properly treating
the combinatorics for $i=j$,
  \begin{eqnarray}
    &&w(N-i-j)\, G_{i,j} 
    \nonumber \\
    &&\longrightarrow  
    w(N-i-j) \Big\lbrack i(1+\delta_{i+1,j}-\delta_{i,j}) G_{i+1,j}
                        \, + j\, G_{i,j+1}\Big\rbrack
                        \nonumber \\
    &&\qquad + w(N+1-i-j)\, G_{i,j}\, .
    \label{B2}
  \end{eqnarray}
The expression (\ref{2.9}) for the two-particle spectrum then 
follows from (\ref{B2}) by complete induction.

\section{Calculation of multiparticle effects}
\label{appa}

Here, we give details of the calculation of higher order Pratt
terms $C_m$ and $G_m$ for the Gaussian source model (\ref{4.1})
in section~\ref{sec4}.
Due to the structure of the functions $f_{ij}$, it is advantageous to
change to relative and average momentum variables. For an
average over $m$ phase space points, we introduce the 
integration variables
  \begin{mathletters}
    \label{A1}
  \begin{eqnarray}
    &&a_i = p_i - p_{i+1}\, ,\, \, 
    b_i = r_i - r_{i+1}\, ,\hbox{for $i \in [1,m-1]$}\, ,
    \label{A1a} \\
    &&a_m = p_m + p_1\, ,\, \, \,
    b_m = r_m + r_1\, .
    \label{A1b}
  \end{eqnarray}
  \end{mathletters}
In terms of these variables, the building blocks for the
$m$-th order Pratt terms read
  \begin{mathletters}
    \label{A2}
  \begin{eqnarray}
    f_{12}f_{23}...f_{(m-1)m} &=& \prod_{j=1}^{m-1}
    {\rm e}^{
      -\textstyle{\sigma^2\over 4}a_j^2
      -\textstyle{1\over 4\sigma^2}b_j^2\, 
      + i\, b_j\, A_j}\, ,
    \label{A2a} \\
    A_n &=& -\textstyle{1\over 2}\sum_{j=1}^{n-1} a_j
            + \textstyle{1\over 2}\sum_{j=n+1}^{m-1} a_j\, ,
    \label{A2b} \\
    f_{m1} &=& 
    {\rm e}^{-\textstyle{\sigma^2\over 4} 
      {\left({\sum_{j=1}^{m-1} a_j}\right)}^2
      -\textstyle{1\over 4\sigma^2}
      {\left({\sum_{j=1}^{m-1} b_j}\right)}^2}
    \nonumber \\
    && \times {\rm e}^{
      - \textstyle{i\over 2}\, a_m\, \sum_{j=1}^{m-1} b_j}\, ,
    \label{A2c} \\
    D_1({\bf P}_1)\, D_m^*({\bf P}_2) &=& 
    {\rm e}^{-\sigma^2 {\left({ a_m/2 - {\bf K}}\right)}^2 
             - i\, {\bf K} \sum_{j=1}^{m-1} b_j}
    \nonumber \\
    &\times& {\rm e}^{-\textstyle{\sigma^2\over 4} 
            {\left({ {\bf q} + \sum_{j=1}^{m-1} a_j  }\right)}^2 
             + i\, {\bf q} \textstyle{b_m\over 2}}\, .
    \label{A2d}
  \end{eqnarray}
  \end{mathletters}
For the Gaussian emission probability (\ref{4.1}), the probability $\rho_2$
of having two particles with wavepackets centered around phase space
points $({\bf r}_i, {\bf p}_i)$, $({\bf r}_j, {\bf p}_j)$, is given
by the product $\rho({\bf r}_i, {\bf p}_i) \cdot$ 
$\rho({\bf r}_j, {\bf p}_j)$ and factorizes into a probability 
distribution for the relative and average particle pair coordinates,
  \begin{mathletters}
    \label{A3}
  \begin{eqnarray}
    \rho_2({\bf p}_i, {\bf p}_j, {\bf r}_i, {\bf r}_j)
    &=& \rho({\bf p}_i, {\bf r}_i)
        \rho({\bf p}_j, {\bf r}_j)
    \nonumber \\
    &=& \rho_{\rm rel}(\textstyle{ {\bf p}_i - {\bf p}_j},
      \textstyle{ {\bf r}_i - {\bf r}_j},)
    \nonumber \\
    && \times \rho_{\rm ave}(\textstyle{ {\bf p}_i + {\bf p}_j},
    \textstyle{ {\bf r}_i + {\bf r}_j})\, ,
    \label{A3a} \\
    \overline{\rho}(a,b) &=& \rho_{\rm rel}(a,b) = \rho_{\rm ave}(a,b) 
    \nonumber \\
    &=& {1\over (2\pi)^3R^3\Delta^3}
    {\rm e}^{-{a^2\over 2\Delta^2}
    -{b^2\over 2 R^2}}\, .
    \label{A3b}
  \end{eqnarray}
  \end{mathletters}
We then define the higher order Pratt terms as averages over
pair distribution probabilities
  \begin{eqnarray}
    G_m({\bf P}_1,{\bf P}_2) &=& 
    \int \left({ \prod_{j=1}^m\, d^3a_j\, d^3b_j\, 
      \overline{\rho}(a_j,b_j)}\right)
    \nonumber\\
  && \times D_1({\bf P}_1)\, f_{12}\, f_{23}\, ...\, f_{(m-1)m}\, 
        D_m^*({\bf P}_2)\, .
    \label{A4}
  \end{eqnarray}
This is a Gaussian integral which can be calculated analytically.
Its exponent is diagonal in all integration variables $b_i$ and
in $a_m$, and doing the corresponding integrals leads to
  \begin{mathletters}
    \label{A5}
  \begin{eqnarray}
    G_m({\bf P}_1,{\bf P}_2) &=& 
    \left( \lbrack 1 + \textstyle{\sigma^2\Delta^2\over 2}\rbrack
    \lbrack1 + \textstyle{R^2\over 2\sigma^2}\rbrack 
    \right)^{\footnotesize{-3(m-1)/2}}\, 
    \gamma^{\footnotesize{-3/2}}
    \nonumber \\
    && \times
    {\rm e}^{-\textstyle{\sigma^2\over 4}
             {\left({ 1 + \textstyle{R^2\over 2\sigma^2}}\right)}
             {\bf q}^2}\, 
    {\rm e}^{-\textstyle{2\over \Delta^2}
             {\left({ 1 - \textstyle{1\over \gamma} }\right)}
             {\bf K}^2}\, 
    \nonumber \\
    && \times I({\bf K},{\bf q})\, ,
    \label{A5a} \\
    \gamma &=& {1\over {1-b}}
    \left({1 + ab(m-1)}\right)\, .
    \label{A5b}
    \end{eqnarray}
  \end{mathletters}
Here, the notational shorthands $a$, $b$ of (\ref{4.3d}) are used again,
  \begin{mathletters}
    \label{A6}
  \begin{eqnarray}
    I({\bf K},{\bf q}) &=& \pi^{-\footnotesize{3(m-1)/ 2}} 
    \int \left({ \prod_{j=1}^{m-1} d^3a_j }\right)
    \nonumber \\
    && \times \exp \lbrack -a_i{\bf M}_{ij}a_j + (w_i+v_i)r_i\rbrack\, ,
    \label{A6a}    \\
    w_i &=& {\bf q}\, \sigma \sqrt{b}\, ,
    \label{A6b} \\
    v_i &=& {\bf K}\, 2\sigma \sqrt{b}a (2i-m)/ \gamma\, ,
    \label{A6c}    \\
    {\bf M}_{ij} &=& \delta_{ij} + b 
    + ab\left({m-1-2{\vert i-j\vert}-\delta_{ij} }\right)
    \nonumber \\
    && - {a^2b^2\over ab(m-1) + 1}\, (2i-m)(2j-m)\, .
    \label{A6d}
  \end{eqnarray}
  \end{mathletters}
The integral $I({\bf K},{\bf q})$ can be calculated explicitly,~\cite{math}
  \begin{mathletters}
    \label{A7}
  \begin{eqnarray}
    \det {\bf M} &=& {h_2^{(m)}\, h_3^{(m)}\over {1 + ab(m-1)}}\, ,
    \label{A7a} \\
    {1\over 4} v_i {\bf M}^{-1}_{ij} w_j &=& 0\, ,
    \label{A7b} \\
    {1\over 4} v_i {\bf M}^{-1}_{ij} v_j &=& 
    2 { {(\gamma -1)h_3^{(m)} - b\gamma h_1^{(m)}}\over 
    \Delta^2 \gamma h_3^{(m)}} {\bf K}^2\, ,
    \label{A7c} \\
    {1\over 4} w_i {\bf M}^{-1}_{ij} w_j &=& 
    {\sigma^2\over 4} {1\over 1-a}
    { {h_2^{(m)} - h_3^{(m)}}\over h_2^{(m)}}\, {\bf q}^2\, .
    \label{A7d}
  \end{eqnarray}
  \end{mathletters}
The factors $h_1^{(m)}$, $h_2^{(m)}$ and $h_3^{(m)}$ are defined in 
(\ref{4.3}) and (\ref{4.4}).


\vskip 1.2cm

\centerline{\bf ERRATUM}

The results presented in section IV are not the results for the
factorizable {\it one-particle} distribution (4.1), in contrast
to our statement. All results 
of section IV were calculated from the probability distribution 
(B3b) for relative and average {\it pair} coordinates according 
to (B4). The use of this pair distribution (B3b) is equivalent 
to the one-particle distribution (4.1) only for the lowest order 
Pratt terms (4.2) on which the conventional two-particle 
symmetrized calculation of the Hanbury-Brown/Twiss effect is based. 
For the higher order Pratt terms, however, our expressions (4.3), 
(4.4) differ from the results for the model (4.1), given in 
Ref. [1]. Especially, all observables of model (4.1) can be shown 
[2] to depend only on the two effective source parameters 
$R_{\rm eff}^2 = R^2 + \sigma^2/2$ and $\Delta_{\rm eff}^2 = 
\Delta^2 + 1/2\, \sigma^2$. The Pratt terms (4.3) and (4.4) on which 
our discussion is based, however, depend on all three 
parameters $R$, $\Delta$ and $\sigma$. This shows that the 
calculation of higher order Pratt terms via pair distributions 
(B4) destroys the one-particle factorizability. We withdraw the
statement that the Pratt terms (4.3), (4.4) are solutions of
the model distribution (4.1). They are solutions of (B4).

\vspace{1cm}

[1] T. Cs\"org\"o and J. Zim\'anyi, Phys. Rev. Lett. {\bf 80} (1998) 916.

[2] U. Heinz, hep-ph/9806512, Conference Proceedings,
    CRIS'98 (Catania, June 8-12, 1998),
    "CRIS'98: Measuring the size of things in 
    the Universe: HBT interferometry and heavy ion physics", 
    (S. Costa et al., eds.), World Scientific, Singapore, 1998. 

\end{document}